# The regulation of symmetry-breaking-dependent electronic structures in ReSeS monolayer

Texture Lin

(School of Physics, Nanjing University, Nanjing 210033, China)

**ABSTRACT:** Due to the excellent physical properties, two dimensional materials have attracted widespread attention from researchers. In this article, we discuss a transition metal dichalcogenide, ReSeS monolayer with $1T''$ phase, with extremely low symmetry through the first principles calculation. It belongs to the space group $P1$ and has Jauns structure. Due to the broken of the central inversion symmetry. Raman and infrared active modes in the vibrational spectrum no longer mutually exclusive, which has become a spectral fingerprint of this material. First principles calculation indicates that the bandgap decreases when the material has geometric deformation, and we propose two models to explain this phenomenon and provide the physical pictures. The first model requires constructing a low symmetry Wannier analytically simplify Hamiltonian to obtain 4 energy levels of mixed orbital, Then the changes of bandgap can be qualitatively described through the difference between these energy levels; The second model analyzes the variations of bandgap through the bonding or antibonding characteristics of the valence and conduction band. We point out that it is a "saturation effect", where the energy of the valence and conduction band has difficulties in further changing under the deformation, resulting in the reduction of the bandgap under both compression and tensile deformation. Meanwhile, we discuss a metallic phase of ReSeS, called $1T'$ phase, to explain the formation of the bandgap. Compared to the $1T''$ phase, it has higher symmetry. Semi-filled d-orbital bands in $1T'$ phases will become unstable and split into bonding and antibonding bands, leading to the opening of bandgap. This process is also known as Peierls phase transition. Finally, we calculated the effective mass of electron and the absorption spectrum. Results point out the potential of ReSeS in the preparation of electrical or optical devices.

E-mail address: texturelin@smail.nju.edu.cn (Texture Lin)

# 目 录

# 第一章　导论

## 1.1 研究背景与内容介绍

2004 年，Novoselov，Geim 与其合作者因成功制备出了少层甚至单层石墨烯，引发了物理学界对低维材料的广泛关注[1]。Geim 团队指出其制备的少层石墨烯即使只有几层原子厚度，却可以在外界环境下表现出很好的稳定性。在室温条件下通过施加一定的栅极电压，多层薄石墨烯体现出大约 $1\times10^4 \text{ cm}^2\cdot\text{V}^{-1}\cdot\text{s}^{-1}$ 数量级的迁移率。其能带结构在薄多层的情况下体现为略微重叠的导带与价带，进而使得整个体系体现为金属性。同时如果改变施加的栅极电压，该体系中电子与空穴对导带与价带的填充情况也会发生相应的变化，进而可以实现对于霍尔系数 $R_H$ 的调节。

事实上由于生长过程中的热涨落，在自然情况下制备出一二维的宏观物质是困难的。虽然如此，Geim 仍指出我们可以通过人工方法制备得到单层石墨烯。其方法包括使用较早的胶带剥离法（scotch-tape technique），与外延生长法。后者是指在其余晶体表面外延生长石墨烯再通过化学蚀刻去除基底以获得少层或单层石墨烯的方法[2]。在单层情况下，石墨烯中电子会表现出更为丰富的物理特性。其中包括：在动量空间高对称 $K$ 点上的狄拉克锥（Dirac cone），这意味着在单层石墨烯晶格中传播的电子完全失去其有效质量，并且需要通过类狄拉克方程对其行为进行描述。同时由于无准经典载流子质量，石墨烯中电子受到的散射较弱，这也使得一些量子特性可以在单层石墨烯中于室温条件下稳定存在，为相关效应的研究提供了较好的平台；其中还包括：由于单层石墨烯仅有一层原子厚度，电子波函数也仅分布在碳原子层附近，其电子波函数对于各类扫描探针与接近原子层的其他材料较为敏感，故而相比于一般的二维电子系统（2D electronic systems），石墨烯存在产生许多新奇现象的可能性。

自单层石墨烯成为可成功制备并稳定存在的第一种二维材料以来，相关二维材料与科研论文，综述呈现出了明显的增长[3-11]。Fig. 1-1 中统计了 2004 到 2015 年间一些二维材料的文章发表数量，该数据也证实了这一趋势。与石墨烯类似，在二维材料中沿着两个方向其原子会呈现出较强的成键效果，进而形成相互作用较强的原子层结构，而在第三个方向一般通过较弱的成键或是分子间作用力相结合，其中包括范德瓦尔斯相互作用（Van Der Waals interaction）。这类材料具有纳

米尺度的厚度与微米甚至毫米数量级的长度和宽度，通常具有很好的电学，光学，力学性能，使得它们在电学器件，光电子学，催化，储能，生物医学，传感器，柔性电子设备等方面有着十分广阔的应用前景[3-6]。举例具体来说：在电学方面，类似于单层石墨烯，二维材料的薄原子层结构允许对其进行载流子，量子与晶格调制。有其参与构建的二维材料同质结（homojuncton）与异质结（heterojunction）也会在界面处产生诸如界面电荷转移，邻近效应（proximity effect）等现象，这些都推动了神经计算（neuromorphic computing）与智能器件的发展。同时二维材料的带隙种类与大小通常受其层数，外界形变，具体结构的影响，而一些二维材料还会具有的特殊的原子轨道填充情况，比如在过渡金属二硫族化合物中过渡金属可能存在半满的 d 轨道电子，这使它们可以在金属性与半导体性之间发生转变。这些都表明了二维材料在光学元件，光电元件设计中具有十分巨大的应用潜力。此外，二维材料具有较低的弯曲刚度与较高的平面内刚度强度，以及较好的延展性，这些还使得二维材料在生物医学器件生产方面起到了关键作用。在制备方面，许多方法已经投入到了二维材料的制作当中。其中包括：机械剥离法（Mechanical exfoliation），液体剥离法（Liquid exfoliation），化学合成法（Chemical synthesis），气体蒸汽生长法（Gas vapor growth）等。通过这些方法可以生产出质量较高的二维材料，并为其各种性能研究提供保障，同时也促进着二维材料的工业化发展。

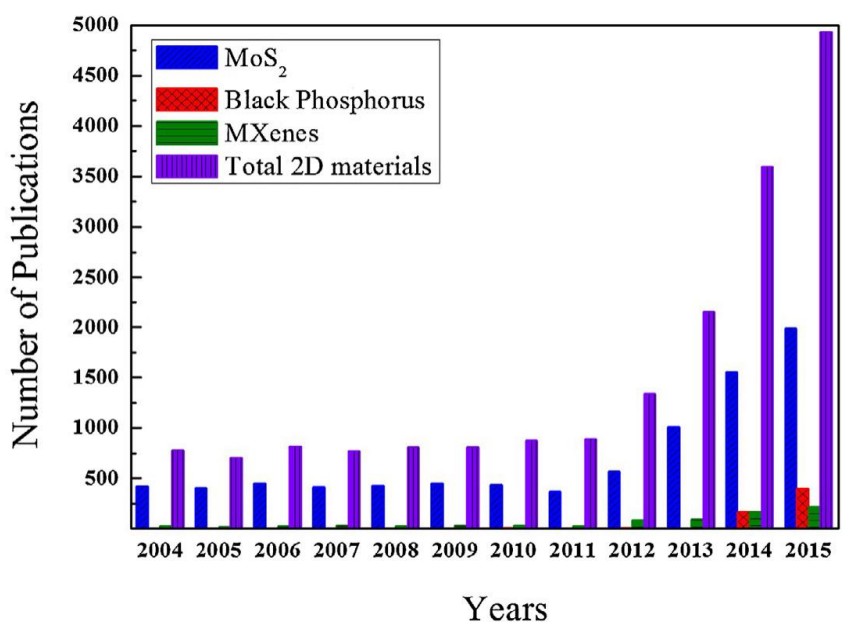

**Fig. 1-1 2004 年到 2015 年相关二维材料文章发表数量统计**[7]

二维材料目前来说种类繁多，如 Fig. 1-2 中所示，我们可以根据其组成成分进行分类。其中一类被称为过渡金属二硫化合物二维材料（two-dimensional transition metal dichalcogenides，2D TMDs），下文简称 TMDs。与石墨烯不同的是，TMDs 多为半导体。这使得 TMDs 有参与晶体管生产的可能性。由于存在一般受层数与几何形变影响的带隙，TMDs 在制造光学调控相关器件方面也具有巨大潜力，多层的结构使得 TMDs 中两层原子层具有较大的比表面积（specific surface area）与间隙，这也使得 TMDs 在电容器制造与传感器方面更具优势。故对 TMDs 性质进行模拟与研究具有重要意义。

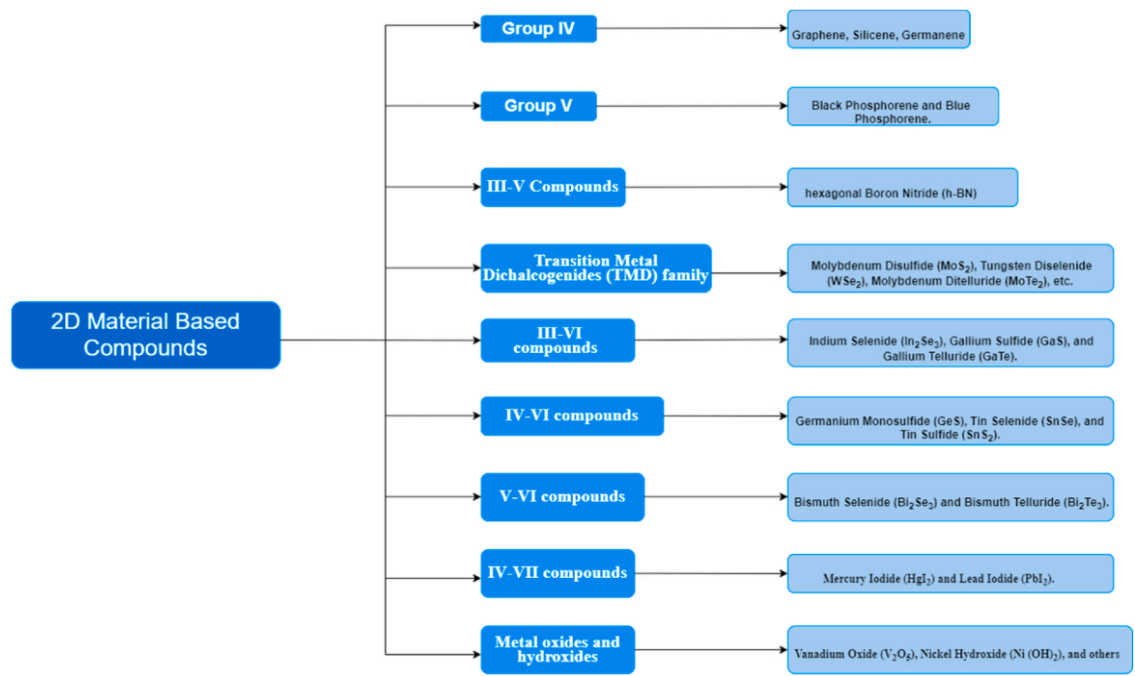

**Fig. 1-2 基于组成成分对二维材料进行分类**[8]

TMDs 的化学表达式一般可以表示为 $MX_2$ 其中 M 表示过渡金属元素而 X 则表示硫族元素。在具体结构中，TMDs 单层通常有三层原子层，金属原子层位于中间，形成了通俗所说的三明治结构。根据金属原子 M 的配位情况，我们可以简单地将 $MX_2$ 分为三棱柱配位（triangular prism coordination）与八面体配位（octahedral coordination）。再根据其在空间中多层堆叠的情况，三棱柱配位可分为 2H 相与 3R 相而八面体配位则有 1T 相。Fig. 1-3 直观的展示了这三种相的堆叠情况。其中 1T 相单层为一个周期，2H 相双层为一个周期，以 ABABAB 方式堆叠，最后 3R 相三层为一个周期，以 ABCABC 方式堆叠[9]。

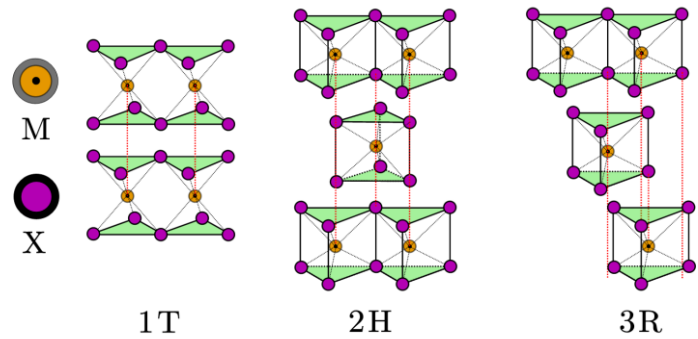

**Fig. 1-3 TMDs 中 MX$_2$ 化合物堆叠情况示意图**

然而 TMDs 可能会在上述结构的基础上发生进一步的畸变以达到稳态，Kertesz 与 Hoffmann 在 20 世纪末期就对三棱柱与八面体配位的 MX$_2$ 进行了研究[12]。结果表明如 ReSe$_2$ 就会发生 Peierls 畸变（Peierls distortion）并使体系能量降低，这类现象也被 Hoffmann 称为固态中的 Jahn-Teller 效应（Jahn-Teller effect）[13]。如 Fig. 1-4(a)中所示，该过程的代表现象是 ReSe$_2$ 中的 Re 原子链会发生 4 聚化，形成所谓的菱形 Re4 链（diamond-shaped Re4 chain）。相关机制之后也得到了研究，其中 Ji-Hae Choi 与 Seung-Hoon Jhi 就在 2018 年就结构与 ReSe$_2$ 完全相同的 ReS$_2$（仅替换元素）进行了理论计算。Jhi 团队指出 ReS$_2$ 除了稳定的结构外，还存在一种亚稳的金属相，为做区分，亚稳态与稳态被分别称为1T′ 相与1T″ 相，Fig. 1-4(b)示意了其两相变化过程中的 4 聚化效应。

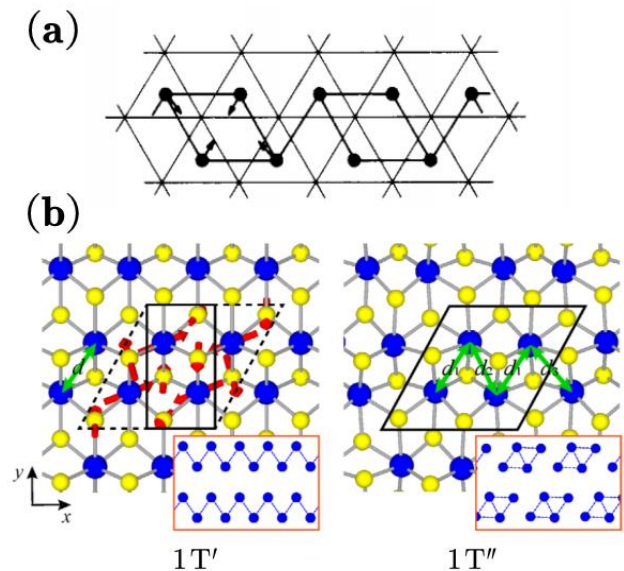

**Fig. 1-4 (a) ReSe$_2$ 发生 4 聚效应的示意图**[12] **(b) ReS$_2$ 由 1T′ 相畸变成为 1T″ 相的过程示意图**[14]

显然，无论是1T′相还是1T″相，体系的对称性都会就1T相发生降低。事实上，1T′相属于 11 号空间群 $P2_1/m$，体系包含螺旋轴操作，镜面反演与中心反演三个对称操作；当1T′相进一步发生对称性破缺而成为1T″相时，便会属于 2 号空间群 $P\bar{1}$，此时该体系仅剩中心反演操作。这些都表明 $ReS_2/Se_2$ 是一种对称性较低的 TMDs 材料。Liu，Fu，Xing 与其团队曾指出 TMDs 会因其晶格具有高度对称性而表现出各向同性行为，而当 TMDs 对称性降低，各类各向异性行为将会被诱导产生。在 $ReS_2$ 中，各向异性导致了其重整化场效应迁移率（renormalized field-effect mobility）沿不同方向具有较大差异，其最大最小值之比高达 3.1，该数值大于大多数被报道的二维材料。同时文章成功的通过加工合成两个 $ReS_2$ EFT 器件得到了一个数字逆变器器件（digital inverter device），体现了低对称性 TMDs 的研究价值与应用潜力[15]。有趣的是，相比于 $ReSe_2$ 与 $ReS_2$，我们还可以更进一步降低对称性，即通过构造 Jauns 结构使得其中心反演对称性发生破缺，并获得 Double-face 的 ReSeS，隶属 1 号空间群 $P1$，除了单位操作外不包含任何对称操作。Fig. 1-5 作为示意图展示了这一过程。具有极低对称性的 ReSeS 的电子结构又具有哪些性质呢？基于这个疑问，我们开展了本篇论文的研究。

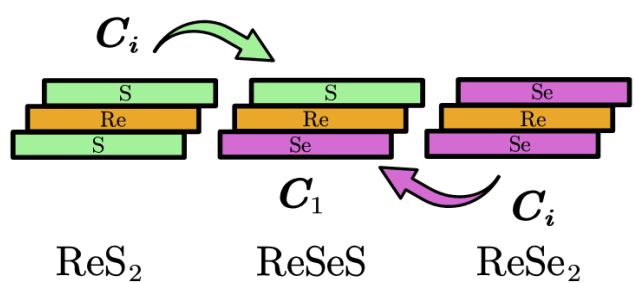

**Fig. 1-5 由 $ReS_2$ 与 $ReSe_2$ 构建 ReSeS 的示意图，其中 $C_i$ 与 $C_1$ 分别表示 $ReS_2/Se_2$ 与 ReSeS 所属的点群**

在这项工作中，我们基于具有 Jauns 结构的单层 ReSeS 进行了第一性原理计算。在正文部分，我们先是在 2.1 节分别介绍了我们使用的两种计算模拟理论。2.1.1 部分我们着重介绍了第一性原理计算的理论基础——密度泛函理论（Density Functional Theory，DFT），2.1.2 部分我们则介绍了一种基于 DFT 能带计算结果的理论方法——最局域万尼尔函数（Maximally Localized Wannier Function，MLWF）理论。之后在 2.2 节中说明了我们进行第一性原理计算时的具体参数设置。

计算的结果与分析在第三章中呈现。3.1 节中我们简要介绍了单层 ReSeS 的基本结构，同时通过分析结构振动模式的红外拉曼活性体现其对称性于单层 $ReS_2/Se_2$ 发生的破缺；3.2 节中我们特别研究了单层 ReSeS 电子结构在几何形变下发生的改变，并提出两个模型，分别在 3.2.1 与 3.2.2 部分中从构建 Wannier 解析简化哈密顿量与原子成键两个角度解释了材料在压缩与拉伸形变下均会缩小带隙的原因。此外，我们在 3.2.2 部分中还讨论了单层 ReSeS 在 1T′ 相转化为 1T″ 相过程具有的晶体场效应图像。3.3 节中我们简要计算了其在倒空间中沿不同方向的电子有效质量，体现了其各向异性，并分析了其在不同几何形变下吸收光谱发生的变化，为观测该材料的形变提供了可能的理论方案。

最后 4.1 节对全文工作进行了统一总结，4.2 节则对未来相关工作进行了展望。

# 第二章　理论背景与方法

## 2.1 使用的理论及其背景

计算物理是一种廉价的研究手段，往往免去了高额的实验仪器费用，却依然可以得到可靠的结果。在本论文中，我们主要使用了两个理论计算方法。分别是 DFT 与 MLWF 理论，在这一节会对它们分别做介绍。DFT 作为第一性原理计算的理论基础，我们会在 2.1.1 部分中简要介绍其发展背景，同时对第一性原理计算思路也进行介绍。MLWF 理论的计算是基于 DFT 计算结果的，二者在计算步骤上有先后关系，在 2.1.2 部分中我们会介绍 MLWF 的理论思路。

### 2.1.1 密度泛函理论（DFT）

一般来说，固体中的电子行为被认为是一种非相对论的多体行为。这也意味着对于固体中电子的动力学行为研究应该求解多体的薛定谔方程（Schrödinger equation）。在定态情况下，我们写出如下的薛定谔方程：

$$\hat{H}\left|\{i\},\{I\}\right\rangle_l = E_l\left|\{i\},\{I\}\right\rangle_l \tag{2-1}$$

在式 2-1 中，多体哈密顿量本征态 $\left|\{i\},\{I\}\right\rangle_l$ 表示有所有电子所在希尔伯特空间中的态 $\left|\{i\}\right\rangle$ 与所有原子核所在希尔伯特空间中的态 $\left|\{I\}\right\rangle$ 张量积的线性组合。这也意味着此时我们不单单考虑固体中的电子行为，原子核也被考虑了进来。这样一个多体定态薛定谔方程无疑是几乎不可能求解的，但我们仍能对其哈密顿量 $\hat{H}$ 进行拆解并得到相关形式。

我们一般将 $\hat{H}$ 拆解为电子之间相互作用项 $\hat{H}_{e-e}$，原子核之间相互作用项 $\hat{H}_{N-N}$ 与电子原子核之间的相互作用项 $\hat{H}_{e-N}$，最后再加上这个系统的外场势能（其余单体能）项 $\hat{V}$。通过轨道-自旋表象进行表示之后，$\left|\{i\},\{I\}\right\rangle_l$ 变成了多体波函数 $\Psi_l(\{q_i\},\{Q_I\})$。其中 $\{q_i\}/\{Q_I\}$ 代表了每个电子与原子核的指标，包括轨道坐标 $\{r_i\}/\{R_I\}$ 与自旋指标 $\{\sigma_i\}/\{\Sigma_I\}$。式 2-2，2-3，2-4 则给出了相应算符的表达形式。在 2.1.1 部分中为讨论方便，除非刻意提起，则默认采用轨道-自旋表象进行表示；为方便记号，记 $4\pi\varepsilon_0 = A$（其中算符标粗是为了提醒算符到函数算子这一变化，一般书籍不做区分）

$$H_{e-e} = \sum_{ij(i \neq j)} \left( \left( \frac{e^2}{2A|r_i - r_j|} \otimes I_{Si} \otimes I_{Si} + \frac{C_{Sij}}{2} \right) \underset{k(k \neq i,j)}{\otimes} I_k \underset{I}{\otimes} I_I \right)$$
$$+ \sum_i \left( -\frac{\hbar^2}{2m_i} \nabla_{r_i}^2 \underset{i' \neq i}{\otimes} I_{Si'} \underset{I}{\otimes} I_{SI} \underset{I}{\otimes} I_I \right) \tag{2-2}$$

$$H_{N-N} = \sum_{IJ(I \neq J)} \left( \frac{B_S(Q_I, Q_J)}{2} \underset{K(K \neq I,J)}{\otimes} I_K \underset{i}{\otimes} I_i \right)$$
$$+ \sum_I \left( -\frac{\hbar^2}{2M_I} \nabla_{R_I}^2 \otimes I_{SI} \underset{I' \neq I}{\otimes} I_{I'} \underset{i}{\otimes} I_i \right) \tag{2-3}$$

$$H_{e-N} = \sum_{iI} \left( A_S(q_i, Q_I) \underset{i'(i' \neq i)}{\otimes} I_{i'} \underset{I'(I' \neq I)}{\otimes} I_{I'} \right) \tag{2-4}$$

$$V = V_e + V_N = \sum_i \left[ V_{Si} \underset{i' \neq i}{\otimes} I_{i'} \underset{I}{\otimes} I_I \right] + \sum_I \left[ \underset{i}{\otimes} I_i \otimes V_{SI} \underset{I' \neq I}{\otimes} I_{I'} \right] \tag{2-5}$$

$I_{Si}$、$I_{SI}$、$I_i$ 与 $I_I$ 是一系列单位算子，表示了电子多体体系中不同希尔伯特空间（Hilbert space）之间的张量积，角标 $I$ 与 $i$ 则分别表示原子核与电子的 Hilbert space。$C_{Sij}$ 表示电子之间的非库伦相互作用。$V_e$ 与 $V_N$ 分别表示电子与原子核的外场势能算子。我们只考虑了单体作用与体系内两体作用，对于一般的体系，这样的考虑已经足够了。但尽管如此，算子形式还是较为复杂，仍然需要进行近似。考虑到原子核的质量是电子的 1000 多倍，我们在考虑电子运动时可以近似原子核是几乎未发生变化的，即原子核仅为电子提供了一个不变的或者说是缓变的势场，而电子绝热于原子核运动。这就是大名鼎鼎的玻恩-奥本海默近似[16]，也称为绝热近似。在这个近似下，多体波函数进行了如下拆分：

$$\Psi_I(\{q_i\},\{Q_I\}) = \chi_I(\{Q_I\}) \otimes \psi_{I\{Q_I\}}(\{q_i\}) \tag{2-6}$$

我们认为其中 $\chi_I(\{Q_I\})$ 仅描述原子核的运动，显然对应于哈密顿量 $H_{N-N} + V_N$，注意此时其中有关电子 Hilbert space 的张量积 $\underset{i}{\otimes} I_i$ 将被平庸化。而 $\psi_{I\{Q_I\}}(\{q_i\})$ 描述了电子的运动，根据绝热近似，此时原子核状态几乎不发生改变，故电子相当于在一个恒定的外场作用中运动，我们再将 $V_e$ 项与 $H_{N-N} + V_N$ 对应的原子核基态能量 $E_{N0}$ 融入其中，并称这个对于电子的全新外场势能算子 $H'_{e-N} = \sum_i V_{Si\{Q_I\}} \underset{i' \neq i}{\otimes} I_{i'}$ 为此时的外场势能算子。于是 $H_e = H_{e-e} + H'_{e-N}$ 就是电子对应的哈密顿量，显然有关原子核 Hilbert space 的张量积 $\underset{I}{\otimes} I_I$ 也被平庸化。当然此

时我们仍然面临一个多体问题，式 2-7 是我们目前拥有的哈密顿量形式。对于固体，一般仅会考虑原子核外的价电子。故而可以将内层紧束缚的电子与原子核看成一个整体，再考察外层电子的运动情况。可以称原子核与其紧束缚的电子为离子实，这个观点不会对方程有任何的简化，只是用离子实代替了原子核，并使得 $H'_{e-N} = \sum_i V_{Si\{Q_l\}} \underset{i' \neq i}{\otimes} I_{i'}$ 的具体形式发生改变。下面简要介绍处理该类多体问题的几种方法。在 2.1.1 部分的讨论中，我们只关心求解这个电子多体系统的基态，这一点在未讨论到 DFT 时不需要特别强调，故我们在对 Hartree 与 Hartree-Fock 近似的介绍中略去了能级指标 $l$。读者也可以认为此时我们一直在求解最低能级指标，即 $l = 0$。

$$H_e = \sum_{ij(i \neq j)} \left( \left( \frac{e^2}{2A|r_i - r_j|} \otimes I_{Si} \otimes I_{Si} + \frac{C_{Sij}}{2} \right) \underset{k(k \neq i,j)}{\otimes} I_k \right) \\ + \sum_i \left[ \left( -\frac{\hbar^2}{2m_i} \nabla^2_{r_i} \otimes I_{Si} + V_{Si\{Q_l\}} \right) \underset{i' \neq i}{\otimes} I_{i'} \right] \tag{2-7}$$

第一种方法被称为 Hartree 近似，其考量极其简单，首先我们不考虑电子与原子核的自旋，仅考虑电子间库仑相互作用，再将电子多体波函数写成多个单体波函数张量积的形式，如式 2-8。最后考虑唯一的约束，即各个单体波函数满足正交归一性：$\left( \varphi_{i\{R_l\}}(r), \varphi_{j\{R_l\}}(r) \right) = \delta_{ij}$。我们使用拉格朗日乘子法对于泛函 $\left( \psi_{\{R_l\}}(\{r_i\}), H_e \psi_{\{R_l\}}(\{r_i\}) \right)$ 进行约束下的极值求解，具体即为对每个单体波函数的共轭 $\varphi^*_{i\{R_l\}}(r)$ 进行变分，得到的结果类似一个单体方程，如式 2-9。其中 $\Lambda_{ij\{R_l\}}$ 是拉格朗日乘子[17]。

$$\psi_{\{R_l\}}(\{r_i\}) = \underset{i}{\otimes} \varphi_{i\{R_l\}}(r_i) \tag{2-8}$$

$$\left( -\frac{\hbar^2}{2m_i} \nabla^2_r + \sum_{j(i \neq j)} \int \frac{e^2 |\varphi_{j\{R_l\}}(r')|^2}{A|r-r'|} dr' + V_{i\{R_l\}} \right) \cdot \varphi_{i\{R_l\}}(r) = \sum_j \Lambda_{ij\{R_l\}} \varphi_{j\{R_l\}}(r) \tag{2-9}$$

Hartree 近似的问题是显然的，除未考虑自旋外，它也没有考虑全同费米子体系的交换反对称性。解决这一问题的一个基础方法就是将电子多体波函数写成单个 Slater 行列式的形式。对于一个电子数为 $n$ 的系统，基态波函数如式 2-10：

$$\psi_{\{Q_I\}}(\{q_i\}) = \frac{1}{\sqrt{n!}} \begin{vmatrix} \varphi_{1\{Q_I\}}(q_1) & \varphi_{1\{Q_I\}}(q_2) & \cdots & \varphi_{1\{Q_I\}}(q_n) \\ \varphi_{2\{Q_I\}}(q_1) & \varphi_{2\{Q_I\}}(q_2) & \cdots & \varphi_{2\{Q_I\}}(q_n) \\ \vdots & \vdots & \ddots & \vdots \\ \varphi_{n\{Q_I\}}(q_1) & \varphi_{n\{Q_I\}}(q_2) & \cdots & \varphi_{n\{Q_I\}}(q_n) \end{vmatrix} \qquad (2\text{-}10)$$

基于这种波函数写法的近似方法被称为 Hartree-Fock 近似，Slater 行列式元素之间此时为张量积运算，其中包含 $n$ 个单体电子波函数。Hartree-Fock 近似延续了 Hartree 近似的思路：首先将这 $n$ 个单体电子波函数的正交归一性 $\left(\varphi_{i\{Q_I\}}(q), \varphi_{i'\{Q_I\}}(q)\right) = \delta_{ii'}$ 作为约束；再通过拉格朗日乘子法对 $\left(\psi_{\{Q_I\}}(\{q_i\}), H_e \psi_{\{Q_I\}}(\{q_i\})\right)$ 中的每个 $\varphi_{i\{Q_I\}}^{\dagger}(q)$ 进行变分，于此同时会得到 $n^2$ 个拉格朗日乘子 $\Lambda_{ij\{Q_I\}}$；最后便能得到 $n$ 个类似单体的微分方程，如式 2-11：

$$\begin{aligned}
&\left(-\frac{\hbar^2}{2m}\nabla_r^2 \otimes I_S + V_{S\{Q_I\}}\right) \cdot \varphi_{i\{Q_I\}}(q) \\
&+ \sum_{i'(i' \neq i)} \left[\left(\left(\int \frac{e^2}{A|r-r'|} \varphi_{i'\{Q_I\}}^{\dagger}(r',\sigma') \cdot \varphi_{i'\{Q_I\}}(r',\sigma')\mathrm{d}r'\right) \otimes I_S\right) \cdot \varphi_{i\{Q_I\}}(q)\right] \\
&- \sum_{i'(i' \neq i)} \left[\left(\left(\int \frac{e^2}{A|r-r'|} \varphi_{i'\{Q_I\}}^{\dagger}(r',\sigma') \cdot \varphi_{i\{Q_I\}}(r',\sigma')\mathrm{d}r'\right) \otimes I_S\right) \cdot \varphi_{i'\{Q_I\}}(q)\right] \qquad (2\text{-}11) \\
&+ C\left[\{\varphi_{i\{Q_I\}}(q)\}\right] \\
&= \sum_{i'} \Lambda_{ii'\{Q_I\}} \varphi_{i'\{Q_I\}}(q)
\end{aligned}$$

式 2-11 中第一项表示单体电子的动能与外场势能；第二项可以看作将其余电子对单体电子波函数的整体平均作用，这两项效应在 Hartree 近似中也存在；第三项则表示电子间的库伦交换作用，与式 2-9 对比可以发现式 2-11 会多出这一项，这是一种量子效应；第四项形式不易具体给出，但其表示了电子间的非库伦相互作用，包括自旋相互作用，也是一种量子效应。

方程 2-9 与 2-11 的求解方法为自洽求解。其简要思路示意图如 Fig. 2-1。核心思想是先给出初始的假设波函数，经过计算得到一组新的波函数，当二者描述的体系差距小于一定标准时认为达到收敛标准。这一过程也被称为自洽循环（self-consistent cycle）。

事实上，如果不考虑自旋且仅考虑电子间库仑相互作用，式 2-11 还可以进一步化简为更为简单熟悉的形式。这里不再做详细描述[18]。但无论如何，式 2-11

中都会包含形式复杂的电子交换作用项，可能还会包含非库伦相互作用项。而 DFT 则通过引入交换关联能 $E_{xc}$ 项巧妙的处理了这一问题，并在很多体系中得到了很好的结果。

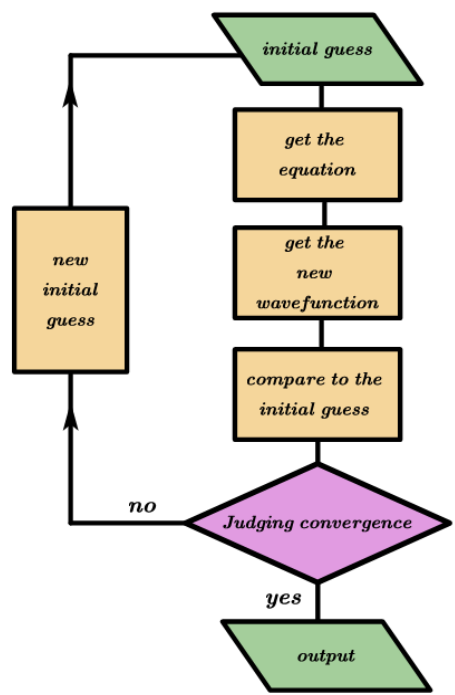

**Fig. 2-1 自洽循环思路示意图**

DFT 指出一个全同的费米子体系能量由三部分构成，分别是体系的动能 $T$，相互作用能 $U$ 与外场势能（其余单体能）$V$。在下面的讨论中，我们始终认为 $T$ 与 $U$ 对应的算子具有恒定不变的形式，仅有 $V$ 对应算子 $V_{Si\{Q_I\}}$ 的形式会发生改变。换句话说，$V_{Si\{Q_I\}}$ 是这个体系中所有物理量的唯一自变量（类似于泛函）。电子多体体系基态的波函数可以写为 $\psi_{0\{Q_I\}}(\{q_i\})$，则其严格的基态粒子数密度为式 2-12，可见 $\rho_0(r)$ 是实空间的函数。

$$\rho_0(r) = \sum_j \left( \int \left| \psi_{0\{Q_I\}}\left(r, \{r_i\}_{i \neq j}, \{\sigma_i\}\right) \right|^2 dr_1 \cdots dr_{j-1} dr_{j+1} \cdots dr_n \right) \tag{2-12}$$

1964 年，Hohenberg 与 Kohn 发表论文并提出了两条重要理论，简称 H-K 定理[19]：

1. 考虑一些基态非简并的不计自旋的全同费米子体系，若它们的外场势能算子是仅以 $r$ 为自变量的函数，则体系的基态外场势能可表达为

$V_0 = \int V_{\{Q_I\}}(r)\rho_0(r)dr$。除去一个平庸的常数，体系的基态粒子数密度 $\rho_0(r)$ 对应于唯一的外场势能算子 $V_{\{Q_I\}}(r)$。

2. 在定理 1 的条件下，粒子数密度是总能 $E(\rho)$ 的唯一变量，且 $E(\rho)$ 对体系基态 $\rho_0(r)$ 取极小值，并等于基态能量。

事实上对一些具体问题，H-K 定理可以进行推广。如 Pan 的团队就在 2015 年将 H-K 定理推广到了在均匀静磁场中运动的电子上，并讨论了考虑电子自旋时的情况[20]。而我们这里仅考虑一种简单的理想情况：

这种理想情况满足两个要求：一是与自旋相关的外场势能算子 $V_{Si\{Q_I\}}$ 可以拆分为 $V_{i\{Q_I\}}(r) \otimes I_{Si} + \Delta_{Si\{Q_I\}}$，且在 $V_{Si\{Q_I\}}$ 的改变中，算子 $\Delta_{Si\{Q_I\}}$ 形式不发生改变，即 $V_{i\{Q_I\}}(r)$ 是体系中所有物理量的唯一自变量，同时 $V_{i\{Q_I\}}(r)$ 是仅以 $r$ 为自变量的函数，还与自旋无关；二是这些体系的基态简并效果仅来源于自旋部分，即体系基态对应唯一的粒子数密度 $\rho_0(r)$。

通过模仿文献[19]中的证明思路，可得到体系基态动能 $T_0$ 与基态相互作用能 $U_0$ 可以表达成仅以 $\rho_0(r)$ 为自变量的泛函，式 2-13 与 2-14 分别给出了形式。忽略任意常数的自由度，基态总能量为式 2-15：

$$T_0(\rho_0) = \sum_j \int \psi_{0\{Q_I\}}^\dagger(\{q_i\}) \left(-\frac{\hbar^2}{2m_j}\nabla_{r_j}^2 \otimes I_{Sj} \bigotimes_{j' \neq j} I_{j'}\right) \psi_{0\{Q_I\}}(\{q_i\}) dr_1 \cdots dr_n \quad (2\text{-}13)$$

$$U_0 = \int \frac{e^2 \rho_0(r)\rho_0(r')}{2A|r-r'|}drdr' + E_{xc}(\rho_0) \quad (2\text{-}14)$$

$$\begin{aligned}E(\rho_0) &= T_0(\rho_0) + U_0(\rho_0) + \int V_{\{Q_I\}}(r)\rho_0(r)dr \\ &\quad + \sum_j \int \psi_{0\{Q_I\}}^\dagger(\{q_i\}) \left(\Delta_{Sj\{Q_I\}} \bigotimes_{j' \neq j} I_{j'}\right) \psi_{0\{Q_I\}}(\{q_i\}) dr_1 \cdots dr_n \quad (2\text{-}15)\\ &= T_0(\rho_0) + U_0(\rho_0) + V_0(\rho_0)\end{aligned}$$

其中 2-13 式被我们写成两项，分别为 Hartree 作用与交换关联作用项。此时 $U_0$ 中并没有如式 2-11 中电子交换作用与非库伦相互作用那样复杂的表达形式，而是用了一个仅以 $\rho_0(r)$ 为自变量的泛函代替。形象地看，可以将 $E_{xc}$ 项当作是一个垃圾箱，我们将体系一些复杂的作用产生的能量放入其中，方便最后统一处理。这也正是上文所提到的：DFT 的巧妙之处。

同时可以看到，由于$\psi_{0\{Q_I\}}(\{q_i\})$具体形式为未知，我们很难在这个基础上对$T_0$与$U_0$再进一步的化简计算。基于这个问题，Kohn 与 Sham 提出了假设（Kohn–Sham conjecture）[21-22]，其核心是将体系基态电子多体波函数写成一个 Slater 行列式的形式代入方程。在这个层面上，DFT 和 Hartree-Fock 近似是类似的，此 Slater 行列式也与式 2-10 具有相同的形式。式 2-16 给出了现在$\rho_0(r)$的表达式。在此基础上，式 2-14 便可以进一步的化简成式 2-17：

$$\rho_0(r) = \sum_i \varphi^\dagger_{0i\{Q_I\}}(r,\sigma) \cdot \varphi_{0i\{Q_I\}}(r,\sigma) \tag{2-16}$$

$$T_0(\rho_0) = \sum_i \int \varphi^\dagger_{0i\{Q_I\}}(r,\sigma) \left( -\frac{\hbar^2}{2m_j} \nabla_r^2 \otimes I_S \right) \varphi_{0i\{Q_I\}}(r,\sigma) \mathrm{d}r \tag{2-17}$$

我们如法炮制：先写出基态总能量$E(\rho_0) = T_0(\rho_0) + U_0(\rho_0) + V_0(\rho_0)$；再通过拉格朗日乘子法对$E(\rho_0)$中的每个$\varphi^\dagger_{0i\{Q_I\}}(q)$进行变分，同时得到$n^2$个拉格朗日乘子$\Lambda_{ij\{Q_I\}}$；最后便能得到$n$个类似单体的微分方程，如式 2-18：

$$\begin{aligned}
&\left[ -\frac{\hbar^2}{2m} \nabla_r^2 \otimes I_S + V_{S\{Q_I\}} + \int \frac{e^2 \rho_0(r')}{A|r-r'|} \mathrm{d}r' + \left( \frac{\delta E_{xc}(\rho)}{\delta \rho} \right)_{\rho=\rho_0} \right] \cdot \varphi_{0i\{Q_I\}}(q) \\
&= \sum_{i'} \Lambda_{ii'\{Q_I\}} \varphi_{0i'\{Q_I\}}(q)
\end{aligned} \tag{2-18}$$

这就是 Kohn-Sham 方程，现在唯一具体形式未知的就是交换关联项$E_{xc}(\rho)$。对于这一项，有一些常用的近似方法，其中包括局域密度近似（Local Density Approximation，LDA）和广义梯度近似（Generalized Gradient Approximation，GGA）。当然对该方程求解也需要自洽循环。如示意图 Fig. 2-1 所示。考虑到比较始末$\rho_0(r)$较为复杂，一般的计算中我们会比较体系的亥姆霍兹自由能，在得到满足收敛标准的自洽解同时，我们还能得到体系的总能量，其中包括离子实之间的相互作用，可以用来分析更多性质。而当我们得到体系的基态后，便可以计算能带，态密度，光学，声子等相关性质，这一步也被称为非自洽循环。

### 2.1.2 最局域万尼尔函数（Maximally Localized Wannier Function）理论

通过自洽循环得到了体系的基态，再通过非自洽循环得到体系能带后，我们还可以通过最局域万尼尔函数（Maximally Localized Wannier Function，MLWF）理论对能带计算结果进行分析。在本文中，这个方法仅仅使用了一点点，故在这

一部分我们不介绍 MLWF 方法的优缺点与细节，而是聚焦于其计算原理。同时由于我们研究单层 ReSeS 体系时未考虑电子自旋。故在 2.1.2 部分的讨论中也不考虑电子自旋。

理想情况下的晶格是无穷的，这也意味着元胞个数 $N$ 是无穷个。能带计算本质上研究的是单电子在周期势场中运动的能级与对应的本征态。通过 DFT 的自洽循环与非自洽循环得到体系的能带后，我们会拥有布洛赫态 $|n,\boldsymbol{k}\rangle$ 与其对应的能量 $E_{nk}$。显然此时体系对应了 4 个好量子数，其中波矢 $\boldsymbol{k}$ 的三个分量分别对应晶格体系中的三个平移算符。同时由于元胞有无穷个，独立的波矢 $\boldsymbol{k}$ 一共也有无穷个，均分布在第一布里渊区（First Brillouin Zone，1BZ）内。对于每一个 $\boldsymbol{k}$，主量子数 $n$ 都可取遍所有主能级，这在 1BZ 上构成了一个类似矢量丛（vector bundle）的代数结构。可以证明，布洛赫态 $|n,\boldsymbol{k}\rangle$ 在倒空间具有周期性，即 $|n,\boldsymbol{k}\rangle = |n,\boldsymbol{k}+\boldsymbol{K}_h\rangle$，其中 $\boldsymbol{K}_h$ 是倒格矢。

不难证明当独立波矢为无穷个时，$e^{i\boldsymbol{k}\cdot\boldsymbol{R}_l}\sqrt{1/v^*}$ 是 1BZ 内一组无穷维正交基，我们用其对 $|n,\boldsymbol{k}\rangle$ 进行展开，即式 2-19，$v^*$ 为 1BZ 的体积。系数 $|w_{n\boldsymbol{R}_l}\rangle$ 满足式 2-20，被称为 Wannier 态，形式上可以看作是 $|n,\boldsymbol{k}\rangle$ 的 Fourier 变换。对于布洛赫态可以进行幺正变换 $\hat{U}$ 以得到一组新的布洛赫态，即：$|n,\boldsymbol{k}\rangle \to \hat{U}|n,\boldsymbol{k}\rangle$。虽然这是规范变换，但却会使得 $|w_{n\boldsymbol{R}_l}\rangle$ 发生改变。我们的基本想法是确定一组量子数 $n$ 相同的布洛赫态 $|n,\boldsymbol{k}\rangle$ 通过傅里叶变换得到确定的一组 $|w_{n\boldsymbol{R}_l}\rangle$，也就是为原始的 $|n,\boldsymbol{k}\rangle$ 找到一个 $\hat{U}$。同时考虑到一般我们只某一能量区域内部的性质，所需要的能带数 $n$ 是有限个的，后面统一为 $M$ 个。故更实际的目标是选定的一个 $\boldsymbol{k}$ 点上的 $M$ 个布洛赫态并通过规范变换 $\hat{U}_{\boldsymbol{k}}$ 构造得到一组新的布洛赫态，并通过这些新的布洛赫态得到 $M$ 组 $|w_{n\boldsymbol{R}_l}\rangle$ 态。此时 $\hat{U}_{\boldsymbol{k}}$ 的矩表示是一个 $M \times M$ 维方阵。

$$|n,\boldsymbol{k}\rangle = \sqrt{\frac{1}{v^*}} \sum_{\boldsymbol{R}_l}^{\infty} |w_{n\boldsymbol{R}_l}\rangle e^{i\boldsymbol{k}\cdot\boldsymbol{R}_l} \tag{2-19}$$

$$|w_{n\boldsymbol{R}_l}\rangle = \sqrt{\frac{1}{v^*}} \int |n,\boldsymbol{k}\rangle e^{-i\boldsymbol{k}\cdot\boldsymbol{R}_l} \mathrm{d}\boldsymbol{k} \tag{2-20}$$

Fig. 2-2 展示了铜的能带结构[23]。观察可以发现该能带结构存在一些能带交叉，如 Fig. 2-2(a)中紫圈所标注的部分。如果以能量范围作为选择能带的判据，

那么很有可能在所选的能量范围内因能带交叉而导致不同 $k$ 点拥有不同维数的 Hillbert space $F_k$，为保证总能选定出 $M$ 个布洛赫态，我们选择的能量范围会大一些，称为能量 *out* 窗口。在该窗口中，若在 $k$ 点的 $F_k$ 为 $M_{outk}$ 维，需要满足 $M_{outk} \geq M$；同时有时我们还想确保在所取的能量范围内一些原始的布洛赫态总能入选，所以还需要设置一个能量 *in* 窗口，每个 $k$ 点位于该窗口内的布洛赫态有 $M_{ink}$ 个，显然满足 $M \geq M_{ink}$。Fig. 2-2(b)展示了三者的关系。

基于以上这些想法，21世纪前后，Nicola Marzari 与 David Vanderbilt 等人开发出了 MLWF 方法[24-25]，并给出了在 *out* 窗口内选择布洛赫态与最终确定规范变换 $\{\hat{U}_k\}$ 的判据。最终的目的就是构造出的 $M$ 组使得 $\Omega$ 最小的 $|w_{nR_l}\rangle$ 态，$\Omega$ 表达式为式 2-21。可以理解为使得 Wannier 函数最局域。式 2-20 中将 $\Omega$ 分成了 $\Omega_I$ 与 $\tilde{\Omega}$ 两部分。可以证明，对于确定的 $|n,k\rangle$（$n$ 取 $M$ 个，$k$ 遍历 1BZ），$\Omega_I$ 对于 $\{\hat{U}_k\}$ 是规范不变的[24]。

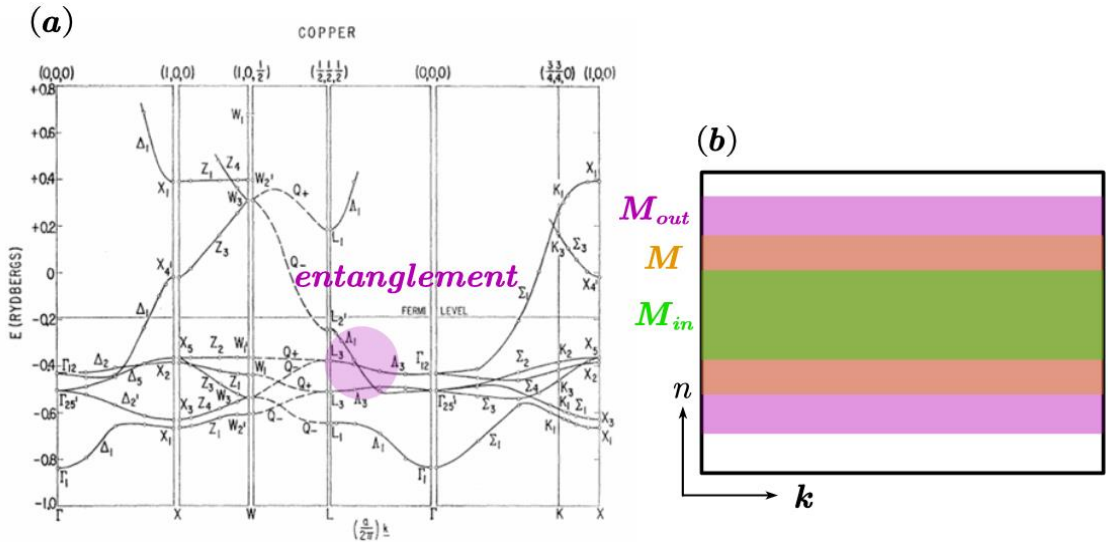

**Fig. 2-2 (a) 晶体铜的能带结构[23] (b) 能量窗口选取示意图**

$$\begin{aligned}\Omega &= \sum_n^M \left( \langle |r|^2 \rangle_{n(R_l=0)} - \langle r \rangle^2_{n(R_l=0)} \right) \\ &= \sum_n^M \left( \langle |r|^2 \rangle_{n(R_l=0)} - \sum_m^M \sum_{R_l} |\langle w_{mR_l}|r|w_{n0}\rangle|^2 \right) + \sum_{mn(m\neq n)} \sum_{R_l(R_l \neq 0)} |\langle w_{mR_l}|r|w_{n0}\rangle|^2 \quad (2\text{-}21)\\ &= \Omega_I + \tilde{\Omega}\end{aligned}$$

在计算机的计算中，倒空间 1BZ 内不可能获得无限个连续的 $k$ 点。一般来说，计算会根据 Monkhorst-Pack 方法在倒空间中取点，我们设此时在 1BZ 中所

取的 $\boldsymbol{k}$ 点有 $N_{kp}$ 个。已知布洛赫态的轨道表示是一个调幅平面波，可以写为 $\psi_{nk} = e^{i\boldsymbol{k}\cdot\boldsymbol{r}} u_{nk}$，其中 $u_{nk}$ 是在实空间中周期为 $\boldsymbol{R}_l$ 的函数，则可以通过定义形式为式 2-22 的交叠式 $G_{mn}^{\boldsymbol{k},\boldsymbol{b}}$ 与有限差分方法写出 $\Omega_I$ 与 $\tilde{\Omega}$ 的离散表达式如式 2-23，2-24。其中 $\boldsymbol{b}$ 表示了倒空间中某一 $\boldsymbol{k}$ 点指向其最近邻所取格点的矢量。$\omega_b$ 是与倒空间中取格点情况相关的权重因子[26]。可证明该离散化之后依然保证了 $\Omega_I$ 的规范不变性。

$$G_{mn}^{\boldsymbol{k},\boldsymbol{b}} = \langle u_{mk} | u_{nk+\boldsymbol{b}} \rangle \tag{2-22}$$

$$\Omega_I = \frac{1}{N_{kp}} \sum_{\boldsymbol{k},\boldsymbol{b}} \omega_b \sum_m^M \left[ 1 - \sum_n^M \left| G_{mn}^{\boldsymbol{k},\boldsymbol{b}} \right|^2 \right] \tag{2-23}$$

$$\begin{aligned}\tilde{\Omega} &= \frac{1}{N_{kp}} \sum_{\boldsymbol{k},\boldsymbol{b}} \omega_b \sum_{n=1}^M \left[ -\text{Im}\left(\ln G_{nn}^{\boldsymbol{k},\boldsymbol{b}}\right) + \boldsymbol{b} \cdot \left( \frac{1}{N_{kp}} \sum_{\boldsymbol{\kappa},\boldsymbol{b}} \omega_b \boldsymbol{b}\, \text{Im}\left(\ln G_{nn}^{\boldsymbol{\kappa},\boldsymbol{b}}\right) \right) \right]^2 \\ &\quad + \frac{1}{N_{kp}} \sum_{\boldsymbol{k},\boldsymbol{b}} \omega_b \sum_{mn(m\neq n)}^M \left| G_{mn}^{\boldsymbol{k},\boldsymbol{b}} \right|^2 \\ &= \Omega_D + \Omega_{OD}\end{aligned} \tag{2-24}$$

考虑到 $\Omega_I$ 对于 $\{\hat{U}_k\}$ 的规范不变性，最小化 $\Omega$ 的过程可分为两步：

1.通过每个 $\boldsymbol{k}$ 点上在 $out$ 窗口内的 $M_{outk}$ 维 Hillbert space $F_k$ 构造出 $M$ 维子空间 $S_k$。使 $\Omega_I$ 最小。考虑到 $M_{ink}$ 中的原始 $|n,\boldsymbol{k}\rangle$ 态一定要入选，其实是在 $M_{outk} - M_{ink}$ 维的 Hillbert space 中找到一个 $M_{outk} - M$ 维的子空间。这一步也被称为解纠缠（Disentanglement）

2.此时已经确定了每个 $\boldsymbol{k}$ 点上的 $S_k$，接下来只用找到 $\{\hat{U}_k\}$ 使 $\tilde{\Omega}$ 最小。

第一步的计算思路依然是在保证 $\langle n,\boldsymbol{k}|n',\boldsymbol{k}'\rangle = \delta_{nn'}\delta_{\boldsymbol{kk'}}$ 的情况下对 $\Omega_I$ 使用拉格朗日乘子法，并对相应的 $\langle n,\boldsymbol{k}|$ 变分，最后再通过自洽求解得到所有 $S_k$。第二步则可以通过最速下降法（steepest-descents）实现。具体的算法已经非常成熟，如本文中使用的 Wannier90 程序包[27]，相关处理细节这里不再叙述。简单总结回顾一下上文：通过对每个 $\boldsymbol{k}$ 点上原始的 $M_{outk}$ 个布洛赫态 $|n,\boldsymbol{k}\rangle$ 构造得到 $M$ 个布洛赫态 $|n,\boldsymbol{k}\rangle_{sub}$，再经过合适的 $\{\hat{U}_k\}$ 得到最终的 $M$ 组 $|w_{n\boldsymbol{R}_l}\rangle$ 态。理想情况下此时的 $|n,\boldsymbol{k}\rangle_{sub}$ 和 $|w_{n\boldsymbol{R}_l}\rangle$ 态之间存在关系如式 2-25，其中 $|n,\boldsymbol{k}\rangle_w$ 为最终得到的布洛赫态。

$$|w_{nR_l}\rangle = \sqrt{\frac{1}{v^*}}\int \left(\hat{U}_k |n,k\rangle_{sub}\right) e^{-ik\cdot R_l}\mathrm{d}k = \sqrt{\frac{1}{v^*}}\int |n,k\rangle_w e^{-ik\cdot R_l}\mathrm{d}k \qquad (2\text{-}25)$$

$M$ 组 $|w_{nR_l}\rangle$ 态在上述条件下构建成功后，一般来说是紧束缚的。这是说每个 $|w_{nR_l}\rangle$ 在实空间对应的波函数都局域在一定的区域。事实上，如果严格按照元胞内原子的轨道数与对应能带数目确定构建的 $|w_{nR_l}\rangle$ 态组数 $M$，并用 DFT 结果中能带对于原子轨道的投影情况（fatband）作为初始子空间猜测（initial guess for the subspaces）[25]，大多数情况下 $|w_{nR_l}\rangle$ 的轨道表示会与孤立的原子轨道十分相似，其波函数中心也会在原子坐标附近。这为我们提供了极好的化学直观。

最后我们简要介绍紧束缚解析模型的推导。我们还是考虑理想情况，且严格按照元胞中的原子轨道数来确定构建的 $|w_{nR_l}\rangle$ 态组数。假设元胞中的原子轨道指标为 $a$，表示某一个原子的某一条轨道，此时计算所得的每组 $|w_{nR_l}\rangle$ 态都会对应一个 $a$，不重不漏。能级指标 $n$ 自然能被我们用原子轨道指标 $a$ 代替，$|w_{nR_l}\rangle$ 写做 $|aR_l\rangle$。我们用这 $M$ 组 $|aR_l\rangle$ 态对能带的单电子哈密顿量 $\hat{H}$ 做展开，如式 2-26。其中 $h_{aa'}^{R_l R_l'}$ 就是我们熟知的 hopping 项，其具体数值可以通过程序如 Wannier90 获得。将 2-25 式带入 2-26 式，化简得到 2-27 式：

$$\begin{aligned}\hat{H} &= \sum_{aR_l}^{\infty}\sum_{a'R_l'}^{\infty}|aR_l\rangle\langle aR_l|\hat{H}|a'R_l'\rangle\langle a'R_l'| \\ &= \sum_{aR_l}^{\infty}\sum_{a'R_l'}^{\infty} h_{aa'}^{R_l R_l'} |aR_l\rangle\langle a'R_l'|\end{aligned} \qquad (2\text{-}26)$$

$$\begin{aligned}\hat{H} = &\int\left(\sum_a h_a |a,k\rangle_w {}_w\langle a,k| + \sum_{aa'(a\neq a')} h_{aa'}|a,k\rangle_w {}_w\langle a',k|\right)\mathrm{d}k \\ &+ \int\left(\sum_{aa'}\sum_{R_l(R_l\neq 0)}^{-\infty\rightarrow\infty} h_{aa'}^{R_l} e^{ik\cdot R_l}|a,k\rangle_w {}_w\langle a',k|\right)\mathrm{d}k\end{aligned} \qquad (2\text{-}27)$$

式 2-27 中前两项分别表示元胞内某一轨道自己的 hopping 与不同轨道间的 hopping，第三项则表示不同元胞间的 hopping。2-27 本质上是一种函数表示展开，因为此时 $k$ 为连续谱。这个形式可以很自然的过渡到二次量子化表示。我们定义轨道电子的生成湮灭算符为 $\hat{a}_k^\dagger$ 与 $\hat{a}_k$。其满足反对易关系 $\{\hat{a}_k^\dagger, \hat{a}'_{k'}\} = \delta_{aa'}\delta_{kk'}$，$\{\hat{a}_k^\dagger, \hat{a}'^\dagger_{k'}\} = 0$ 与 $\{\hat{a}_k, \hat{a}'_{k'}\} = 0$。式 2-28 改写为 2-29。此时，我们定义出 $\hat{H}$ 的核

（Hamiltonia kernel）$\mathbf{H}_k$ 如式 2-29。其维数与原子轨道数相同，在每个 $k$ 点均可解得 $M$ 个本征值，即该 $k$ 点对应的能带能量 $E_{nk}$。

$$\hat{H} = \int \left( \sum_a h_a \hat{a}_k^\dagger \hat{a}_k + \sum_{aa'(a \neq a')} h_{aa'} \hat{a}_k^\dagger \hat{a}_k' + \sum_{aa'} \sum_{R_l(R_l \neq 0)}^{-\infty \to \infty} h_{aa'}^{R_l} e^{ik \cdot R_l} \hat{a}_k^\dagger \hat{a}_k' \right) \mathrm{d}k \quad (2\text{-}28)$$

$$\mathbf{H}_k = \begin{pmatrix} h_1 + \sum\limits_{R_l(R_l \neq 0)}^{-\infty \to \infty} h_{11}^{R_l} e^{ik \cdot R_l} & h_{12} + \sum\limits_{R_l(R_l \neq 0)}^{-\infty \to \infty} h_{12}^{R_l} e^{ik \cdot R_l} & \cdots & h_{1M} + \sum\limits_{R_l(R_l \neq 0)}^{-\infty \to \infty} h_{1M}^{R_l} e^{ik \cdot R_l} \\ h_{12}^\dagger + \sum\limits_{R_l(R_l \neq 0)}^{-\infty \to \infty} h_{12}^{R_l\,\dagger} e^{-ik \cdot R_l} & \ddots & \ddots & \vdots \\ \vdots & \ddots & \ddots & \vdots \\ h_{1M}^\dagger + \sum\limits_{R_l(R_l \neq 0)}^{-\infty \to \infty} h_{1M}^{R_l\,\dagger} e^{-ik \cdot R_l} & \cdots & \cdots & h_M + \sum\limits_{R_l(R_l \neq 0)}^{-\infty \to \infty} h_{MM}^{R_l} e^{ik \cdot R_l} \end{pmatrix} \quad (2\text{-}29)$$

应该指出，$\mathbf{H}_k$ 的维数取决于展开 $\hat{H}$ 的 $|w_{nR_l}\rangle$ 组数。我们容易验证式 2-24 中的 $|w_{nR_l}\rangle$ 态相互正交，但由于一般研究中我们关注的能带为有限个，所构造的 $|w_{nR_l}\rangle$ 组数也为有限个，故这些 $|w_{nR_l}\rangle$ 态仅满足子空间的归一性。重要的是，在解纠缠步骤中，我们在 *out* 窗口中构造出了维度小于等于 $F_k$ 的子空间 $S_k$，这意味着部分能带信息会发生缺失；而 *in* 窗口中所有的原始布洛赫态 $|n,k\rangle$ 都原封不动的光荣入选第二步，和最终的 $|n,k\rangle_w$ 仅差一个规范变换，这又意味着这些 $|n,k\rangle$ 对应的能带信息一定会被保存。故 *in* 窗口也被称为 *frozen* 窗口。

还应该指出，我们在上述分析时将 $|w_{nR_l}\rangle$ 态的能级指标 $n$ 用原子轨道指标 $a$ 进行了替换，读者可能会认为 $|w_{nR_l}\rangle$ 态的求解结果必然像极了位于元胞中的孤立原子轨道。事实上，在某些情况下，一个 $|w_{nR_l}\rangle$ 态也可以很像许多原子轨道的组合，即混合轨道。这是因为其本质仅仅是提取 DFT 结果中的布洛赫波信息并做 Fourier 变换，并不天生具有表示轨道的物理意义，该点在 3.2.1 部分中还会进一步讨论。

## 2.2 参数设置

这一小节我们主要介绍本论文采用的具体参数设置。本文主要进行的是 DFT 计算，计算对象是单层 ReSeS。所使用的程序为 Materials Studio 中的 CASTEP[28] 模块与 VASP[29-30] 程序包。同时针对 VASP 的计算结果，采用了 Wannier90，WannSymm[31]，与 Lobster[32-33] 程序进行分析。下面分别介绍它们的具体应用。

我们首先从材料网站 Materials Project[34]上面获得了空间群为 $P\bar{1}$ 的 $ReS_2$ 多层结构，在 Materials Studio 中对其进行切片得到单层 $ReS_2$，其中真空层设置大于$15\mathring{A}$。之后再对其替换原子得到单层 ReSeS 与单层 $ReSe_2$，并进行收敛性测试。相关结构的截断能与 Monkhorst-Pack $\boldsymbol{k}$ 格点，我们统一在表 2-1 中展示。

在 CASTEP 模块中参数设置如下：

1.对单层 $ReSe_2$ 与单层 $ReS_2$ 我们仅计算了振动光谱与能带结构，振动光谱包含拉曼与红外振动模式。由于振动光谱涉及到结构的受力，故在结构优化中我们采用了较高的精度。CASTEP 模块中结构优化的精度会从 4 个维度进行评估：结构的能量，单位为 $eV/atom$；结构中所受的最大力，单位为 $eV/\mathring{A}$；结构中所受的最大应力，单位为 $GPa$；结构中的最大位移，单位为 $\mathring{A}$。我们设置的精度分别为 $5.0\times10^{-7}\ eV/atom$，$0.005\ eV/\mathring{A}$，$0.01\ GPa$，$5.0\times10^{-4}\ \mathring{A}$。这已经超过了软件推荐的最高标准。对于自洽循环，我们采用的自洽循环收敛标准（SCF tolerance）为 $1.0\times10^{-7}\ eV/atom$。

2.对于单层 ReSeS 的1T′/T″ 相，我们在计算电子性质的同时，还计算了声子谱。CASTEP 模块在计算声子谱时对结构精度有更高要求。计算 1T′ 相的电子性质时，优化 4 参数为 $5.0\times10^{-6}\ eV/atom$，$0.01\ eV/\mathring{A}$，$0.02\ GPa$，$5.0\times10^{-4}\ \mathring{A}$。尽管略微降低了精度，这依然是软件推荐的最高标准。而1T″ 相计算电子性质时的优化精度则与单层 $ReS_2/Se_2$ 相同；计算声子谱时1T′/T″ 相均采用了参数：$1.0\times10^{-6}\ eV/atom$，$0.005\ eV/\mathring{A}$，$0.005\ GPa$，$5.0\times10^{-5}\ \mathring{A}$。所有情况下 SCF tolerance 标准均为 $1.0\times10^{-7}\ eV/atom$。

3.特别指出，由于单层 ReSeS 为 Jauns 结构，我们对其施加了偶极修正。在 CASTEP 模块中这会导致在优化晶胞时报错。故我们将1T″ 相电子性质计算的优化分为了两步：首先优化晶胞-不施加修正；其次不优化晶胞-施加偶极修正，且在后续计算中施加偶极修正。在1T″ 相声子谱计算的优化中我们全程未施加偶极修正，而在计算声子谱时施加偶极修正。（原因是初期认为偶极修正可能导致声子谱出现虚频，后期证实这在 CASTEP 模块中影响不大，故未能保持一致，但这并不影响本文对声子谱分析得到的结论，计算结果依旧可靠）。

相比于 CASTEP 模块，VASP 参数设置较为简单，我们也未用其计算声子相关性质。对于单层 ReSeS 的1T′/T″ 相，结构优化标准均为 $0.02\ eV/\mathring{A}$，SCF tolerance 标准均为 $1.0\times10^{-5}\ eV$。对于1T″ 相优化依然采用：首先优化晶胞-不施加修

正；其次不优化晶胞-施加偶极修正的方法。考虑到我们仅计算了能带等简单的电子性质，以上精度已经足够。

下面来说明截断能与 Monkhorst-Pack $k$ 格点的参数设置，这是 DFT 中最为重要的参数之一，对其设置的本质目的是为了平衡计算资源与计算精度，故需要对其进行收敛性测试。我们统一以亥姆霍兹自由能为判断标准：对于截断能，半导体体系每提高100 eV，单个原子能量变化不超过1.0 meV，金属性体系变化不超过1.2 meV；对于 Monkhorst-Pack $k$ 格点，每个方向约提高取点数 1，体系单个原子能量变化不超过1.0 meV。表 2-1 罗列了相关结构最终所采用的截断能与 Monkhorst-Pack $k$ 格点设置。在 VASP 中，我们使用的是 PAW 赝势；在 CASTEP 中，我们使用了 OTFG 模守恒（OTFG norm conserving）赝势。对于交换关联泛函，二者均选择了 GGA-PBE。

表 2-1 相关结构最终确定的截断能与 Monkhorst-Pack $k$ 格点设置

| 结构 | 截断能 | Monkhorst-Pack $k$ 格点 |
| --- | --- | --- |
| VASP-1T″ReSeS | 500 eV | $9 \times 9 \times 1$ |
| VASP-1T′ReSeS | 500 eV | $18 \times 9 \times 1$ |
| CASTEP-1T″ReSeS | 1450 eV | $4 \times 4 \times 1$ |
| CASTEP-1T′ReSeS | 1550 eV | $11 \times 4 \times 1$ |
| CASTEP-$ReS_2$ | 900 eV | $4 \times 4 \times 1$ |
| CASTEP-$ReSe_2$ | 1500 eV | $4 \times 4 \times 1$ |

Wannier90，WannSymm，与 Lobster 程序作为分析工具，这里不再详细介绍。其中 Wannier90 的原理在 2.1.2 部分中已经做过说明，WannSymm 作为一种对称化工具，可以对 Wannier90 结果进行处理并给出符合结构对称性的解析哈密顿量，式 2-29 已经给出了解析哈密顿量的核的表达式，而本文中所有 Wannier90 的计算结果均通过 WannSymm 进行了对称化。Lobster 则是一款成熟的分子成键分析工具，能给出体系不同能量范围内总的或特定原子之间的成键与反键情况，这对我们在 3.2.2 部分中分析能带性质有很大帮助。

最后我们特别向读者强调本文晶格形变的实现方式，相比于一般实验上的单轴应变，我们在改变一轴长度的同时限制了另一轴的改变。这也就意味着我们施加的本质上是一种双轴应变，并称其为几何形变。具体操作流程是将元胞某轴拉伸/压缩到一定程度，再固定元胞对原子进行结构优化。对不同几何形变下的结构，我们采用 CASTEP 进行计算，参数设置与未形变的单层 ReSeS 保持一致。

# 第三章 单层 ReSeS 的电子结构与形变调控

这一部分我们将展开对单层 ReSeS 电子结构的讨论，具体内容分为三节。3.1 节作为介绍部分，关注的是结构的晶格参数，对称性与基本电子性质，3.2 节讨论的重点是电子结构的改变，包括1T′/T″两相转变过程中发生的改变，与在几何形变下发生的改变。3.3 节则是对和应用相关性质的计算。

## 3.1 单层 ReSeS 的基本性质

ReSeS 是 TMDs 材料中的一种，体块中有明显的层状结构，层与层之间通过较弱的分子间作用力相结合，使得单层 ReSeS 成为了一种二维材料。单层 ReSeS 中仍然具有三层原子，Se/S 原子层位于上下两侧，这也意味着单层 ReSeS 具有 Jauns 结构，且对比于结构相似的单层 $ReS_2/Se_2$ 具有更低的对称性。ReSeS 元胞中含有互不相同的 12 个原子，稳定结构属于 $P1$ 空间群，这意味着它不具有任何非平庸的点群对称操作！ReSeS 中的金属原子是 8 配位的，故属于 TMDs 家族中的1T 大类；与 $ReS_2/Se_2$ 相似，其稳定结构晶格会发生较为明显的畸变，因此被称为1T″ 相，Fig. 3-1(a)展示了 $ReS_2$，ReSeS 与 $ReSe_2$ 的晶格结构，他们的晶格常数列在了表 3-1 中。其中除了晶格夹角 $\theta$ 变化较小之外，晶格矢 OA 与 OB 皆会随着 Se 原子在化合物中的占比增高而扩大，这是很好理解的：Se 原子相比于 S 原子拥有更大的原子半径，与 Re 原子形成的共价键会有更长的键长，进而导致了元胞晶格的扩张。

表 3-1 单层 $ReS_2$ ReSeS $ReSe_2$ 的晶格常数

| 结构 | OA$\left(\overset{\circ}{A}\right)$ | OB$\left(\overset{\circ}{A}\right)$ | $\theta(°)$ |
| --- | --- | --- | --- |
| $ReS_2$ | 6.51 | 6.40 | 118.85 |
| ReSeS | 6.65 | 6.53 | 118.84 |
| $ReSe_2$ | 6.78 | 6.66 | 118.82 |

Fig. 3-1(b)是此类 TMDs 的 1BZ。由于晶格矢 OA 与 OB 十分接近，其元胞格子有类似于石墨烯的六角结构，进而其 1BZ 是一个近正六边形。但同时，它们的对称性远低于石墨烯，结构也具有更强的各向异性，故 1BZ 中具有不等价的三组 M 与 K 点。图中标注出的 Γ 点与所有 M，K 点共同组成了该布里渊区的不可约表示，该区间能够反映出电子的能谱信息。

对比 ReSeS 与 $ReS_2/Se_2$ 结构，可以直观地感受到 Jauns 结构带来的对称性破缺：Double-face 的特点使得 ReSeS 不再具有中心反演对称性。同时这也意味着上下面的电子结构将不再对称，静电势可以很好的体现这一点。如 Fig. 3-1(c)所示，在晶体中原子核或离子实分布的区域，静电势数值上会出现一些极小值，说明电子在这些区域趋向于稳定分布。与 $ReS_2/Se_2$ 对称的静电势结构不同，ReSeS 因 Se，S 原子的核电荷数差异会在二者的原子层之间形成 7.5 eV 左右的势能差。相比于 S 原子，Se 原子更大的核电荷数会使周围分布的电子具有更低的能量，在空间中沿着 $z$ 轴形成一个类似于一维势阱的结构。

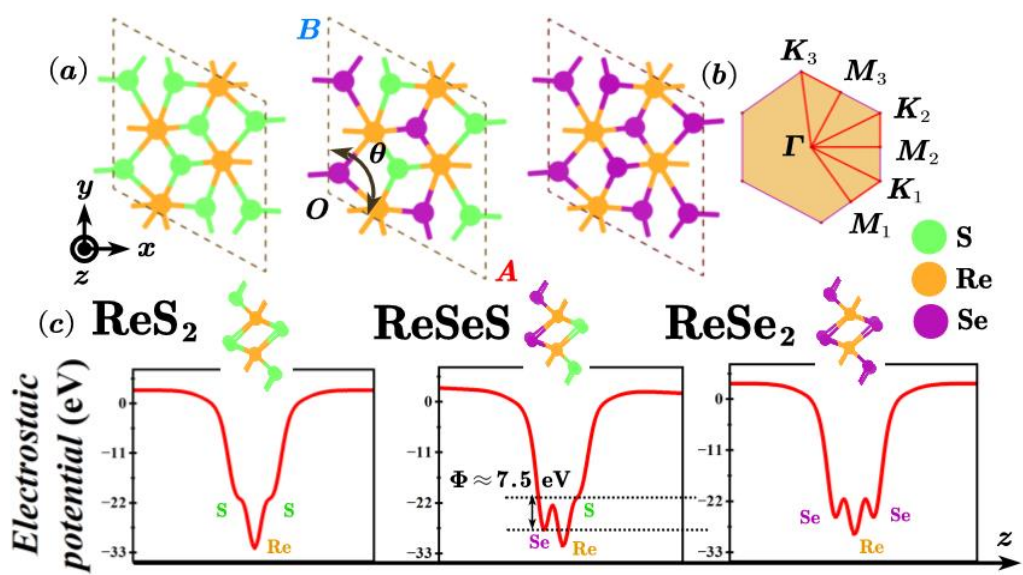

**Fig. 3-1 单层 $ReS_2$ ReSeS $ReSe_2$ 的 (a) 晶体结构，(b) 第一布里渊区(1BZ)和 (c) 静电势分布**

在影响电子分布的同时，中心反演的对称性破缺也会造成振动光谱发生改变。分子的振动光谱可以基本分为红外吸收光谱（infrared absorption spectrum）与拉曼光谱（Raman spectrum）。一般来说，若分子的振动改变了体系的偶极矩，该振动模式可能具有红外活性；相对应的，改变了体系的极化率的振动模式可能具有拉曼活性。通过群论可以证明，对于一个中心对称的分子，其具有红外活性的振动模式与具有拉曼活性的振动模式满足互斥关系。对于单层 $ReS_2/Se_2$，其元胞中均含有 12 个原子，声子谱会形成 36 条色散关系，其中包含代表整体运动的 3 条声学支与 33 条光学支；同时其结构皆属于点群 $C_i$，显然包括中心反演操作，由 $C_i$ 群的不可约表示可知振动具有 $A_g$ $A_u$ 两种模式。事实上，$ReS_2/Se_2$ 二者均拥有 18 个具有拉曼活性的振动模式与 15 个具有红外活性的振动模式。而对于毫

无对称性的单层 ReSeS，其不可约表示仅有 A 一种，除声学支之外的振动模式都同时具有红外与拉曼活性，这是其作为 Jauns 结构所具有的本征特性，进而可以作为单层 ReSeS 的指纹特征用于材料的区分鉴别和遴选，具体频率在表 3-2 中罗列。

表 3-2 单层 $ReS_2$ ReSeS $ReSe_2$ 振动光谱对应频率及其拉曼红外活性

| $ReS_2$ | | ReSeS | | $ReSe_2$ | |
|---|---|---|---|---|---|
| 频率 ($cm^{-1}$) | 模式 | 频率 ($cm^{-1}$) | 模式 | 频率 ($cm^{-1}$) | 模式 |
| 130.30 | $A_g$ | 111.00 | A | 100.87 | $A_g$ |
| 134.98 | $A_u$ | 117.56 | A | 113.97 | $A_u$ |
| 137.78 | $A_g$ | 119.35 | A | 116.66 | $A_g$ |
| 148.93 | $A_g$ | 123.38 | A | 117.45 | $A_g$ |
| 154.55 | $A_u$ | 129.55 | A | 119.16 | $A_u$ |
| 159.36 | $A_g$ | 141.27 | A | 123.89 | $A_g$ |
| 210.21 | $A_g$ | 163.95 | A | 139.07 | $A_g$ |
| 216.52 | $A_u$ | 169.86 | A | 144.14 | $A_u$ |
| 230.70 | $A_g$ | 185.53 | A | 166.54 | $A_g$ |
| 259.70 | $A_u$ | 191.62 | A | 171.59 | $A_u$ |
| 262.91 | $A_g$ | 195.88 | A | 172.75 | $A_g$ |
| 263.24 | $A_u$ | 199.21 | A | 173.49 | $A_u$ |
| 269.64 | $A_g$ | 206.65 | A | 180.89 | $A_g$ |
| 275.62 | $A_u$ | 208.22 | A | 184.74 | $A_u$ |
| 296.74 | $A_g$ | 223.15 | A | 188.20 | $A_g$ |
| 297.13 | $A_u$ | 224.27 | A | 191.49 | $A_u$ |
| 299.06 | $A_g$ | 229.29 | A | 194.26 | $A_g$ |
| 300.91 | $A_u$ | 240.69 | A | 198.59 | $A_u$ |
| 304.29 | $A_g$ | 252.08 | A | 201.57 | $A_g$ |
| 311.91 | $A_g$ | 254.18 | A | 207.37 | $A_u$ |
| 312.77 | $A_u$ | 264.96 | A | 209.32 | $A_g$ |
| 332.18 | $A_u$ | 271.34 | A | 215.28 | $A_u$ |
| 332.86 | $A_g$ | 278.19 | A | 228.36 | $A_g$ |

| ReS₂ | | ReSeS | | ReSe₂ | |
|---|---|---|---|---|---|
| 频率(cm⁻¹) | 模式 | 频率(cm⁻¹) | 模式 | 频率(cm⁻¹) | 模式 |
| 351.53 | $A_u$ | 283.56 | A | 228.93 | $A_u$ |
| 355.04 | $A_g$ | 292.77 | A | 231.70 | $A_g$ |
| 362.53 | $A_u$ | 300.55 | A | 242.06 | $A_u$ |
| 363.09 | $A_g$ | 301.54 | A | 244.39 | $A_g$ |
| 383.41 | $A_u$ | 317.67 | A | 245.23 | $A_u$ |
| 393.39 | $A_g$ | 338.86 | A | 260.25 | $A_g$ |
| 406.19 | $A_g$ | 350.37 | A | 296.20 | $A_g$ |
| 411.12 | $A_u$ | 375.18 | A | 300.36 | $A_u$ |
| 424.25 | $A_g$ | 397.39 | A | 303.79 | $A_g$ |
| 443.62 | $A_u$ | 422.90 | A | 320.88 | $A_u$ |

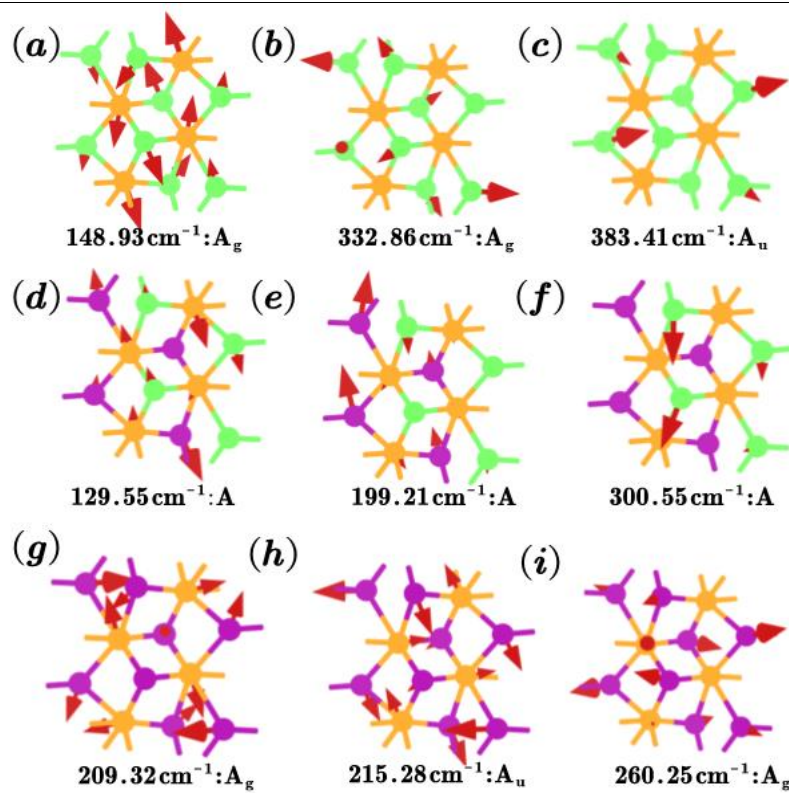

**Fig. 3-2** 单层 (a)-(c) ReS₂ (d)-(f) ReSeS (g)-(i) ReSe₂ 的相关振动模式。红色箭头为振动方向，下标中 ：前为频率，：后为模式

与此同时，我们挑选了一些典型振动模式进行可视化。如 Fig. 3-2(a)-(c)与(g)-(i)所示，$A_g$ 模式很明显呈现出中心对称的振动形式而 $A_u$ 模式则表现为中心非对

称。相比之下，Fig. 3-2(d)-(e)中 ReSeS 则可以粗略看作是以上两种振动模式的叠加混合。横向对比三列模式还可以发现在低频区域 Re，Se/S 均有发生振动，而随着频率升高，往往仅有体系中的硫族元素能够保持振动。这可以通过谐振机制简单理解，频率 $v$ 正比于 $\sqrt{k/m}$，在成键的劲度系数一定时，质量轻的原子更容易发生高频的振动。而在 TMDs 中，金属一般具有较大的相对原子质量，进而金属的振动主要分布在低频区域。当然这一点从表 3-2 的数据中也能反映出来：不难发现随着 Se 元素在体系中的占比提高，所有模式的频率整体呈现出下降趋势。

## 3.2 单层 ReSeS 电子结构在形变下的改变

3.1 节中我们简要讨论了单层 ReSeS 的基本结构，电子性质与振动光谱。其目的仅仅是为了反应 ReSeS 相较于 $ReS_2/Se_2$ 所发生的对称性破缺。而在这一节中，我们将从基本的电子色散关系出发，分析 ReSeS 的电子结构在几何形变下发生的变化。

和大多数 TMDs 一样，ReSeS 在稳态 1T″ 相下是半导体。如 Fig. 3-3(a)所示，其具有大约 1.3 eV 的带隙，能带路径中包含有三个不等价的三角回路，每个三角回路近似为六角晶格的不可约布里渊区。其价带顶（VBM）并不严格位于 Γ 点处，能带色散在该高对称点附近会表现出微小的起伏。考虑到其导带底（CBM）严格位于 Γ 点处，我们称其带隙种类为近直接带隙（near-direct bandgap）。相似的现象在 $ReS_2/Se_2$ 中也能观察到。比较三种结构的带隙大小，可以发现随着 Se 原子在结构中的比例上升，带隙 $E_g$ 呈现出缩小的趋势（$ReS_2/Se_2$ 带隙分别为 1.43 eV 与 1.21 eV），这说明 Se 掺杂可能能够加强该体系的金属性，而相关研究的也印证了这一点[36]。

对于基本的电子结构，功函数（work function）可以简单粗略地反应费米面电子能量与真空能级能量之差，这三类材料的功函数列举在表 3-3 中。考虑到对于半导体材料，VBM 与费米面具有完全相同的物理意义，费米能级 $\varepsilon_f$ 与真空能级 $E_{vac}$ 之差既可以认为是功函数的取值，也可以认为是 VBM 的能量值，而 CBM 能量便可以通过关系 " $\varepsilon_f - E_{vac} + E_g$ " 求出。由于 ReSeS 非对称的 Double-face 结构，功函数在 S Se 原子层两侧并不相同，故在表格中我们还计算了平均效应。正如 3.1 节中所讨论的，考虑到 Se 原子更大的核电荷数，从 Se 一侧溢出电子显然需要消耗更多能量，进而伴随有更大的功函数。同时 Se 原子相比于 S 具有能级更

高的外层电子,这意味着体系内的电子能够占据能量更高的能态,故 $ReSe_2$ 拥有更高的 V/CBM 能量。结合 Fig. 3-3(b)-(c)与表 3-3,仔细观察 V/CBM 的变化率可以发现:价带与导带的色散关系是相似的,但是在单层 $ReSe_2$ 的 V/CBM 能量上升的过程中,VBM 上升更加迅速,导致了二者能量差缩小,$ReSe_2$ 也因此具有了更小的带隙。这说明,影响 $ReSe_2$ 带隙缩小的主要原因并不是某支能带具体色散关系的改变,而是不同能带间能量变化程度的差异。这种带隙改变的机制是这类材料的一个重要特点,在后面的分析中可以看到,该特点也可以解释在几何形变下 ReSeS 带隙所发生的变化。

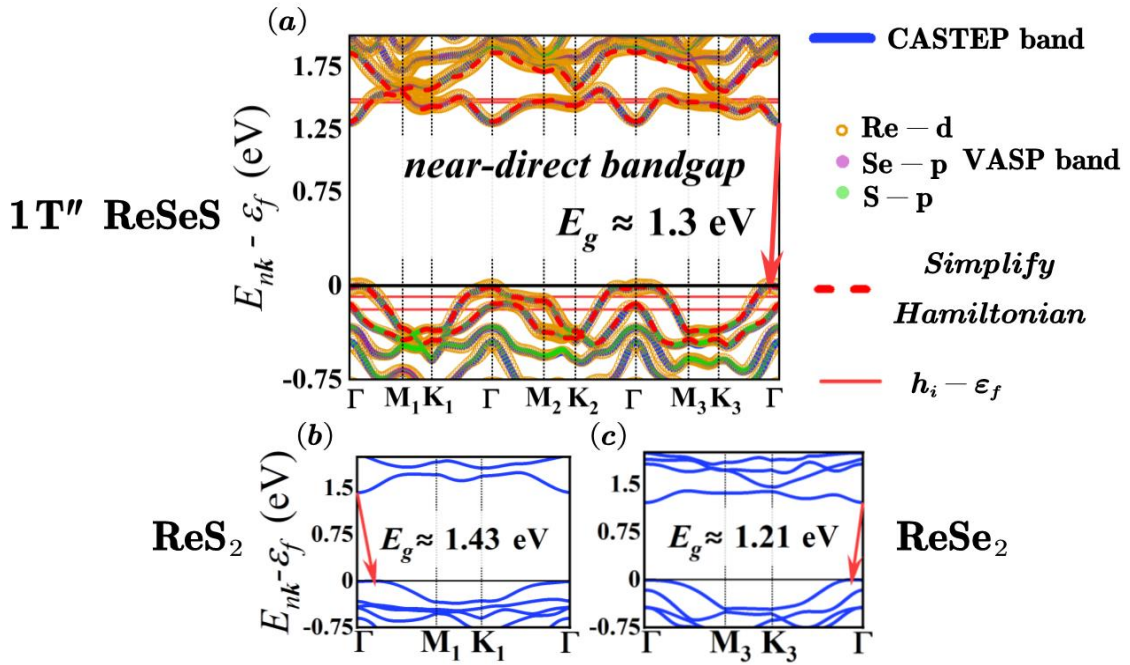

**Fig. 3-3 (a)** 单层 1T″ 相 ReSeS 的能带,蓝色实线为 CASTEP 所得能带,黄紫绿三色气泡大小代表由 VASP 求出的能带对相应原子轨道的投影大小,红色虚线为由 Wannier 构造的简化哈密顿量求出的能带,红色实心直线代表简化哈密顿量核的对角元与费米能级之差 **(b)** 单层 $ReS_2$ 的能带 **(c)** 单层 $ReSe_2$ 的能带

表 3-3 单层 $ReS_2$ ReSeS $ReSe_2$ 的功函数

|  | $ReS_2$ | ReSeS | | | $ReSe_2$ |
|---|---|---|---|---|---|
|  |  | S - side | Se - side | Average |  |
| $\varepsilon_f - E_{vac}$ (VBM) | -5.68 eV | -4.79 eV | -5.87 eV | -5.33 eV | -4.89 eV |
| $\varepsilon_f - E_{vac} + E_g$ (CBM) | -4.25 eV | -3.49 eV | -4.57 eV | -4.03 eV | -3.68 eV |

二维材料通常具有良好的力学敏感性,意味着其光学,电学性质很容易通过外部应力进行调控[37-38]。而应力的施加也通常意味着材料的几何形变,在本篇论

文中，考虑到ReSeS的低对称性，我们沿其不同方向施加几何形变以探究其能带结构的改变。我们定义了形变因子$\delta_i$，表示结构在$i$方向发生的形变大小。具体数值满足关系$\delta_i = (l'_i - l_i)/l_i$，$l'_i$与$l_i$分别表示形变前后$i$方向的晶格常数。ReSeS具有两个明显各向异性的元胞基矢方向——OA OB 轴方向，我们计算了其在不同轴长下的能带数值，结果如 Fig. 3-4(a)所示。整体来看，无论沿着什么方向的拉伸压缩形变都会使得ReSeS带隙呈现出减小的趋势。同时相较于OA 轴方向的变化，OB 轴形变对带隙的调控作用显然更加明显。这是由于1T″相 ReSeS 中 Re 原子链方向（OB 方向）的相互作用强于链间方向（OA 方向），进而沿着 OB 方向的几何形变会对能带能量造成相对显著的影响。

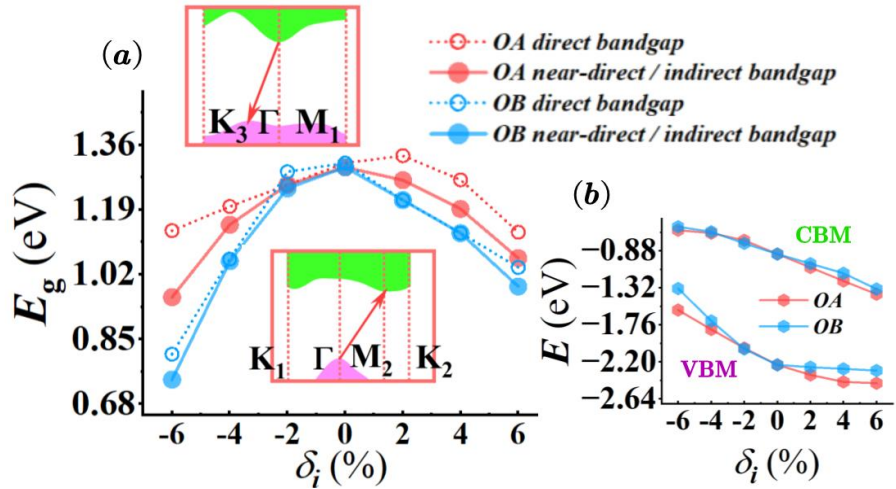

**Fig. 3-4 (a) 1T″相 ReSeS 带隙在不同方向几何形变下的改变，红蓝分别代表沿 OA，OB 轴，实线虚线分别代表了直接与近直接/间接带隙。红色小方框中紫色代表价带，绿色代表导带**
**(b) ReSeS 能带 V/CBM 在不同方向几何形变下的改变，红蓝分别代表沿 OA，OB 轴，能量高的自然为 CBM，能量低的为 VBM。**

带隙的改变本身仅取决于 VBM 与 CBM 的变化，但影响其变化的机制可以是多种多样的。初步来看，导带或价带不同$k$点能量变化可能会导致带隙改变，包括其大小与种类。Fig. 3-4(a)中通过示意图描述了 ReSeS 中两种主要的由能带色散形状改变所引起的带隙改变机制：当Γ点附近价带能量值起伏较大时，如(a)中上小图，带隙大小将改变但种类仍然能保持为近直接带隙；而当在倒空间 1BZ 边界处的导带能量发生大幅下降时，如(a)中下小图所示，此时不仅带隙大小发生改变，其种类也会变成真正意义上的间接带隙。不过无论那种变化机制，都会使得近直接/间接带隙与直接带隙差值增大，而在我们计算的形变区间内，色散的

变化程度与材料发生的几何形变成程度正相关，这说明了在大形变下不同种类带隙差异将更加明显，Fig. 3-4(a)中实虚线在不同 $\delta_i$ 处取值也反映出了这一点。

上述的分析并没有完全涵盖 ReSeS 带隙变化的机理。事实上，在计算中，导带与价带内部的色散关系改变涉及到的能量尺度通常为 0.05 eV 到 0.1 eV，偶尔可以达到 0.2 eV。这已经可以使带隙发生较为明显的变化了，但相较于 Fig. 3-4(a) 中的计算结果，该能量显然并没有完全达到形变所造成的带隙减少。Fig. 3-4(b) 中不同形变下 V/CBM 的能量也说明了这一问题：仅靠单一能带内部不同 $\boldsymbol{k}$ 点能量起伏，难以达到 V/CBM 的能量变化尺度。这说明除了能带具体色散形状的变化，能带的整体能量改变也是导致带隙缩小的重要机制，下面在 3.2.1 与 3.2.2 部分将对此进行详细的讨论。

### 3.2.1 解析简化哈密顿量模型

对于孤立体系，例如位于真空中的各类分子，其能级量子数通常是离散化的。而对于含有较多元胞的晶体，空间平移对称性使得其在倒空间中形成了连续的能谱。对于这个过程，我们可以看作是孤立分子能级的相互作用导致了能级劈裂，进而形成了有一定展宽的能带。相反的，每支能带的能量也可以反映出对应能级的能量大小。对于 ReSeS，Fig. 3-3(a)中通过 VASP 实现的轨道投影能带说明了在费米面附近 Re 原子 d 轨道与 S，Se 原子的 p 轨道存在交叠。故对于一个元胞内的分子能级，也应该是 d 轨道与 p 轨道能级通过相互作用共同形成的结果。每个 Re 原子包含 5 条 d 轨道，每个 S，Se 原子均拥有 3 条 p 轨道，理论上元胞内的成键电子涉及到 44 个能级，对应 44 条能带。由于 ReSeS 体系不存在任何对称性方面的限制，这是十分复杂的。所以我们尝试对其构造简化哈密顿量（simplify Hamiltonian）模型以直观体现其能级能量变化。仿照式 2-28 与 2-29，其哈密顿量核的解析形式如式 3-1 所示：

$$\mathbf{H}_{\boldsymbol{k}} = \bigoplus_{i=1}^{4} \boldsymbol{h}_i + \boldsymbol{\Delta}_{\boldsymbol{k}}$$

$$= \begin{pmatrix} h_1 & 0 & 0 & 0 \\ 0 & h_2 & 0 & 0 \\ 0 & 0 & h_3 & 0 \\ 0 & 0 & 0 & h_4 \end{pmatrix} + \begin{pmatrix} \Delta_{11}(\boldsymbol{k}) & \Delta_{12}(\boldsymbol{k}) & \cdots & \Delta_{14}(\boldsymbol{k}) \\ \Delta_{12}^{\dagger}(\boldsymbol{k}) & \ddots & \ddots & \vdots \\ \vdots & \ddots & \ddots & \Delta_{34}(\boldsymbol{k}) \\ \Delta_{14}^{\dagger}(\boldsymbol{k}) & \cdots & \Delta_{34}^{\dagger}(\boldsymbol{k}) & \Delta_{11}(\boldsymbol{k}) \end{pmatrix} \quad (3-1)$$

其中 $\boldsymbol{h}_i$ 是单轨道作用的哈密顿量核。而 $\boldsymbol{\Delta}_{\boldsymbol{k}}$ 代表了 $\boldsymbol{h}_i$ 对应的四个单轨道之间的相互作用，其代表的矩阵均为厄密矩阵。求解出的能带用红色虚线展示在 Fig.3-3(a)中，其中的四条红色实心直线表示了 $h_i - \varepsilon_f$，$\varepsilon_f$ 为费米能级。可以看到该模型与 DFT 能带有较好吻合，而 $h_i - \varepsilon_f$ 的能量范围也与费米面附近的四条能带能量相近。此时 $h_i$ 可以看作是混合轨道的能级，我们可以认为 4 个无色散关系的能级在相互作用下形成了具有色散关系的能带，所有形变下的 $h_i$ 数值统计在了表 3-4 中。

表 3-4 单层 1T″ 相 ReSeS 不同形变下简化哈密顿量核中 $h_i$ 的取值与方差

|  | $h_1$ (eV) | $h_2$ (eV) | $h_3$ (eV) | $h_4$ (eV) | $S_\delta$ (eV) |
|---|---|---|---|---|---|
| 原结构 | -2.43 | -2.32 | -0.77 | -0.75 | 0.65 |
| $\delta_{OA}$ = -6% | -2.08 | -1.99 | -0.45 | -0.33 | 0.68 |
| $\delta_{OA}$ = -4% | -2.21 | -1.50 | -1.25 | -0.40 | 0.42 |
| $\delta_{OA}$ = -2% | -2.21 | -2.11 | -0.85 | -0.65 | 0.51 |
| $\delta_{OA}$ = 2% | -2.53 | -2.44 | -0.93 | -0.73 | 0.69 |
| $\delta_{OA}$ = 4% | -2.54 | -2.53 | -0.98 | -0.93 | 0.62 |
| $\delta_{OA}$ = 6% | -2.54 | -2.50 | -1.16 | -1.10 | 0.48 |
| $\delta_{OB}$ = -6% | -1.50 | -1.26 | -1.10 | -0.94 | 0.04 |
| $\delta_{OB}$ = -4% | -2.15 | -0.49 | -0.60 | -0.49 | 0.49 |
| $\delta_{OB}$ = -2% | -2.34 | -2.78 | -0.63 | -0.55 | 0.74 |
| $\delta_{OB}$ = 2% | -2.13 | -2.04 | -1.33 | -1.21 | 0.17 |
| $\delta_{OB}$ = 4% | -2.19 | -1.75 | -1.73 | -1.41 | 0.08 |
| $\delta_{OB}$ = 6% | -2.66 | -1.84 | -1.50 | -1.45 | 0.23 |

应该指出，此时 $h_i$ 所代表的能级的物理意义是模糊的，Wannier90 需要通过原子轨道做初始子空间猜测，而在 ReSeS 中我们使用了 Re 原子的 s 轨道作为初始猜测，进而才能够得到 4 组 Wannier 态。但实际能带中费米面附近并不具有明显的 s 轨道特性，这也意味着这样构造出的 Wannier 函数难以局域化且不能精确描述能带特性，只能看作是一种近似，故相应哈密顿量被称作为简化哈密顿量。在式 2-29 中，原则上应该考虑无穷个元胞之间的 hopping。但当构造所得的 Wannier 函数具有良好的局域性时，我们仅用考虑近邻乃至次近邻之间的 hopping。由于以上构造的简化哈密顿量的 Wannier 函数局域性很差，故我们不得不考虑更多元胞之间的 hopping 以得到较好的计算结果。具体考虑的 hopping 关系如 Fig. 3-5 所示，可以看到一共考虑了约 4 层近邻元胞之间的 hopping。

Wannier90 在计算中对于 hopping 项没有明确的对称性限制，故求出 hopping 值可能并不严格满足体系对称性。但是 ReSeS 结构本身不具有对称性，而 hopping 项本质上是由体系提供的 DFT 结果求出，没有引入任何额外参数，这又说明对于 ReSeS，只要能求出 hopping 项，便可以毫无限制的构建该类简化模型。显然，对于对称性好的体系，简化模型通常会因为其物理意义过于粗糙而难以构建。而尽管 ReSeS 可以构建出质量尚可的简化哈密顿量，我们也只能用其结果定性分析。我们将此时 Wannier 态对应的轨道称为混合轨道，4 条混合轨道能级如果本身具有相近的能量值，它们在 $\Delta_k$ 项的作用下形成的能带也会具有较小的能量差，对应于较小的带隙。为简单衡量 $h_i$ 的能量差异，我们计算了其方差 $s_\delta$，结果为表 3-4 的最后一列。

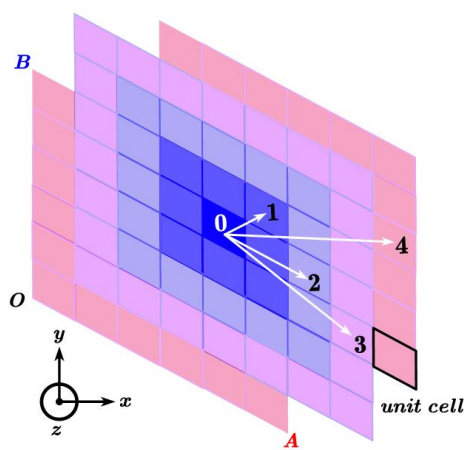

**Fig. 3-5 简化哈密顿量所考虑的不同元胞间的 hopping 的示意图，其中数字从 0 到 4 依次代表：从元胞内的 hopping 到元胞与第 4 层近邻之间的 hopping**

不难看出，当材料发生了几何形变，方差大多数表现为减小，意味着 4 个混合轨道能级的靠近与较小的带隙，这与 Fig.3-4(a)中所展现出的规律相符。同时随着形变由压缩变为拉伸，$h_i$ 的数值整体呈现出减小的趋势，这又符合 Fig.3-4(b) 中 V/CBM 发生的改变。总而言之，这个粗糙的模型为我们提供了一种并不严谨的直观图像，分析 $h_i$ 的过程中忽略了不同 $k$ 点色散对能带的影响，但反映出了影响带隙变化的另一因素：能带能量的整体改变。不过它并没有回答能带整体能量改变的具体原因，相关机制还需要我们进一步探索。

### 3.2.2 成键反键能级模型

为了更好回答能带整体能量变化的机制，我们需要先由能级出发讨论带隙的形成，对于 8 配位的 ReSeS，其稳定结构显然发生了严重的畸变。而从正六角格

子转变为4聚体的过程中，还存在一个亚稳的金属相，称为1T′相，这是我们理解能隙来源的关键。Fig. 3-6(a)简要对比了1T′相与1T″相ReSeS的结构差异，1T′相属于6号空间群$P_m$，仅具有镜面反演操作。由于对称性的上升，其元胞仅含6个原子。类似于1.1节中提到的$ReS_2$两相，Re原子链在1T′相转变为1T″相的过程中由zig-zag行变为4聚化的diamond形状。

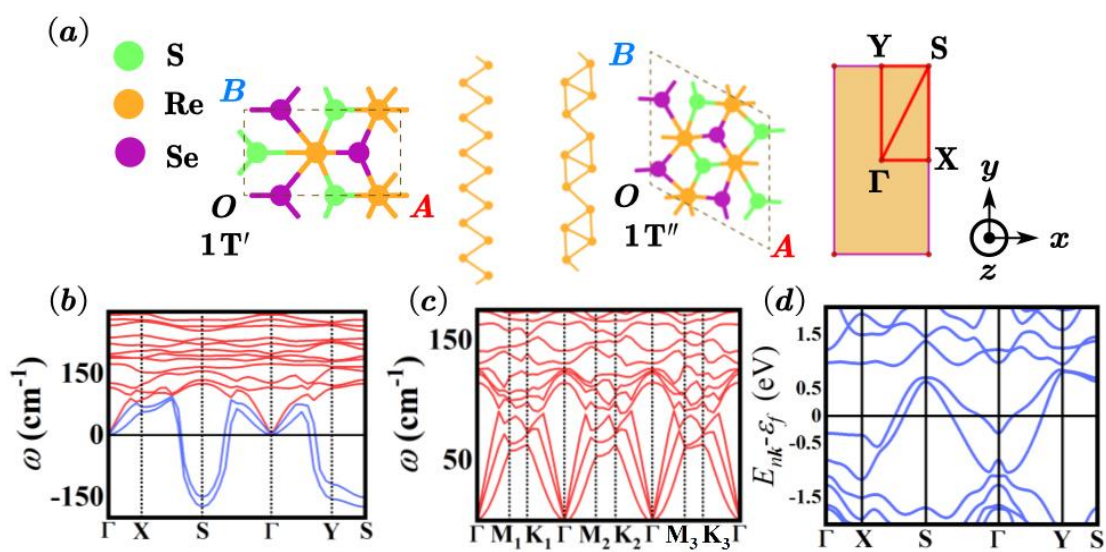

**Fig. 3-6 (a) 1T′/1T″相单层ReSeS的结构与1T′相单层ReSeS的1BZ (b)-(c) 1T′/1T″相单层ReSeS的声子谱，蓝色代表含虚频的声学支 (d) 1T′相单层ReSeS的能带结构**

Fig. 3-6(b)-(c)通过声子谱的计算反映出了1T′相的亚稳定性——两只声学支在高对称S点附近出现了较大虚频。1T′相的能带结构沿倒空间$y$轴具有较大起伏而在$x$方向变化平缓，体现出了类似一维单原子链的色散特性[39]，这说明1T′相与1T″相类似：链内（OB方向）的Re原子之间存在更强的相互作用。

无论是1T′相还是1T″相，ReSeS都体现出8配位的晶体场特点。以1T″相为例，Fig. 3-7(a)展示了其中Re原子与周围6个S，Se原子形成的畸变8面体。根据晶体场理论，d轨道会根据受到配体排斥作用的强弱而发生劈裂，沿8面体轴向的两个轨道因受到的排斥作用强而升高，一般记为$e_g$；其余三个轨道相应的降低能量使体系趋于稳定，称为$t_{2g}$，而其轨道波函数主要分布在轴向间隙。

ReSeS中，元胞中的4个Re原子能够提供20个d轨道，对应20条能带。Se与S原子均有空余的外层价电子轨道，可以认为Re原子外层的s轨道电子与部分d轨道电子会对其填充,而填充后的每个Re原子将仅剩3个d轨道价电子。根据外层电子数量，可知费米面以下仅需 6 个轨道便可填充满所有 d 轨道电子

（考虑 Pauli 不相容），故剩余的 14 个轨道全部对应于导带。下面统一将能量最低的能带称为 1 号能带，考虑到 8 配位晶体场中 $e_g$ 轨道能量较高，我们认为 13 到 20 号左右的能带的电子波函数会具有 $e_g$ 轨道的特点。Fig. 3-7(b)绘制了一些能带对应的轨道作为示例，如图，1T″相中的 7 号能带所对应的电子轨道波函数主要分布在 Re-Se，Re-S 键间，这正是 $t_{2g}$ 轨道的明显特征；而 17 号能带的电子轨道则沿着成键方向分布，也正代表了 $e_g$ 轨道的特点。对于1T′相而言，其元胞中原子数量减半，对应轨道也自然需要减半处理，图中展示了其 4 号/10 号能带对应的 $t_{2g}/e_g$ 轨道特点的电子波函数。

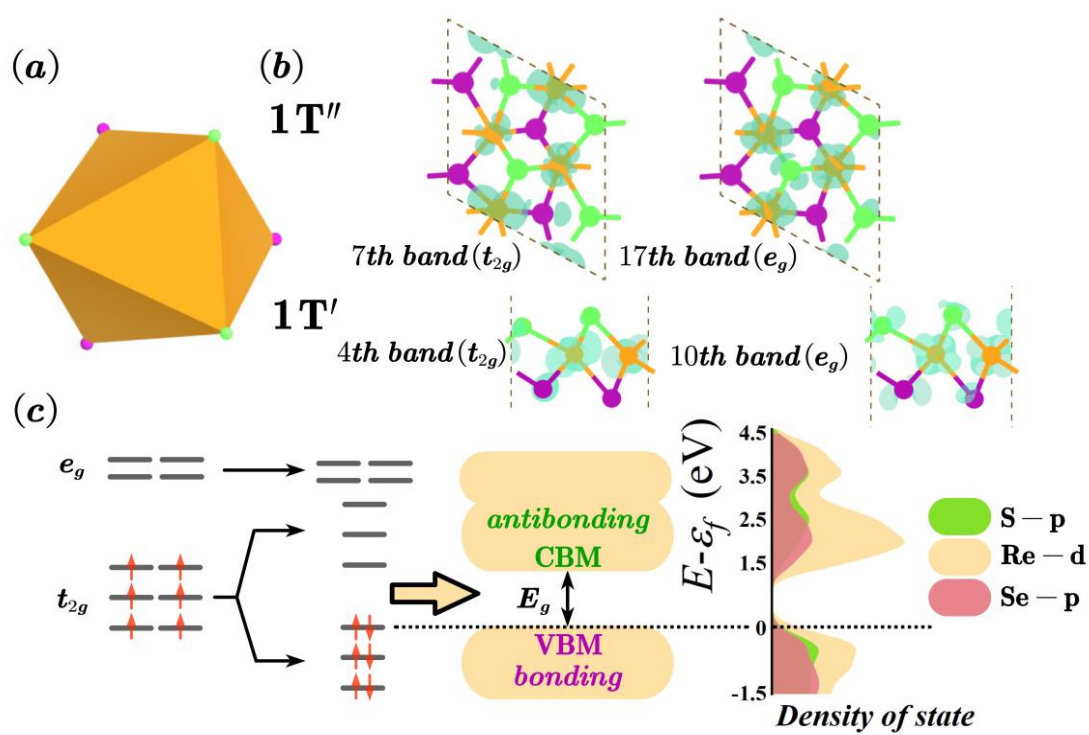

**Fig. 3-7 (a) 1T″相单层 ReSeS 的畸变八面体配位 (b) 1T′/1T″相单层 ReSeS 不同能带所对应的电子轨道，数字代表能带序号，括号内表示轨道的种类 (c) 1T′/1T″相转变过程中能级变化的示意图，最右侧为 DFT 计算出的1T″相单层 ReSeS 的投影态密度，不同颜色代表不同的原子轨道**

与1T″相不同的是，1T′是金属相。这意味着费米能级会穿过其能带，导致电子的半填充。若简单地用能级表示，可以理解为 6 个 $t_{2g}$ d 轨道各自仅填充一个电子。这种填充方式很可能导致体系能量失稳，即偶数个半满的轨道发生劈裂，形成能量较高的反键轨道与能量较低的成键轨道，而电子则全部填充到低能成键轨道上使整个体系的电子能量降低，并打开一个带隙导致结构由金属相变为半导

体相，这就是所谓的 Peierls 相变，也是1T′/T″相之间的转变的重要原理[40]。其示意图如 Fig. 3-7(c)所示，从中可以看到分析的图像与 DFT 计算所得的态密度具有很好的对应。此外，Fig. 3-8(a)还展示了我们通过 Lobster 程序计算出的1T′/T″相中 Re-Re 原子间成键的 COHP 值，COHP 结果为负代表成键，正值代表反键。结果表明 Re-Re 键在1T″相中的费米能级上下呈现出明显的成键反键差异，也与我们上述的分析相符。这些都说明了 Re-Re 轨道成键-反键劈裂就是促使1T′/T″相变发生并形成带隙的主要机制。

不过需要注意的是，电子能量的降低并不意味着这种转变一定能够自然发生，事实上，伴随着轨道劈裂，晶体结构往往也会跟着改变，这可能导致晶格弹性势能上升，故还需要比较相变前后总能量的高低。DFT 计算可以得到体系的亥姆霍兹自由能 $F$，有关系 $F = E - TS$，其中 $E$ 是体系的内能，$T$, $S$ 分别是体系的温度与熵。我们先比较 0 温时的情况，由于1T′相元胞原子是1T″相的二分之一，我们对其阔胞计算。结果得出1T′相能量高出1T″相约 1.2 eV，印证了该相变可以自然发生的事实。当然如果系统具有一定温度，1T′相的部分电子可能会被激发到更高能级，进而在 Peierls 相变的过程中电子能量下降的幅度将会减小，甚至部分电子可能因热激发而占据能量更高的反键态。此时电子总能下降可能低于晶格弹性势能的增加，故高温并不利于该类畸变发生。

在上述的分析里我们认为相变的过程中 Re-Re 之间的成键反键发挥着主要作用，虽然这能很好的解释带隙形成的原因并提供清晰的物理图像，但其并不能完整的反映出1T″相 ReSeS 中的成键关系。对于价带与导带，Re 原子轨道与 S, Se 原子的 p 轨道也会发生交叠，研究整个体系的成键自然应该包含这一部分。我们使用 Lobster 程序计算了1T′/T″相结构内所有原子对成键的 COHP 值，结果如 Fig. 3-8(b)所示。可以看到相比于 Re-Re 成键，ReSeS 在价带附近整体呈现出强度较小的反键，这一特性在我们施加的形变范围内均未改变，Fig. 3-7(c)-(f)中展示了相关数据。

V/CBM 的成键与反键特性对于应力调控具有重要意义，Singh 的团队就曾指出反键轨道所导致的能量下降会使得磷烯在一定应力下由半导体变为金属[41]，而对于1T″相 ReSeS 我们也能提出类似的模型：由于其价带与导带整体都呈现出反键特性，当我们拉伸材料，相关反键作用将会减弱进而导致 VBM 与 CBM 的能量下降；相对的，压缩材料会使 V/CBM 的能量上升，这也符合 Fig. 3-4(b)计

算所得的结论。而无论是拉伸形变还是压缩形变，ReSeS 的带隙都有减小的趋势，这是由 V/CBM 的反键强弱差异所导致的，我们也可以将其看作是一种"饱和效应"。Fig. 3-8(b)说明了导带的反键强度明显大于价带，当材料拉伸，价带与导带的反键都将减弱，然而由于价带本身就只有十分微弱的反键效果，随形变程度其反键强度很快达到饱和，即能量随应变难以再减小，故 CBM 更大的下降幅度导致了带隙的缩小；而当材料发生压缩形变时，导带的强反键效果也会趋于达到饱和，相比之下，VBM 的能量快速上升，也同样的导致了带隙的缩小。也就是说，这两种相对的趋势促成了相同的结果：即拉伸与压缩形变均可以使 ReSeS 带隙缩小。

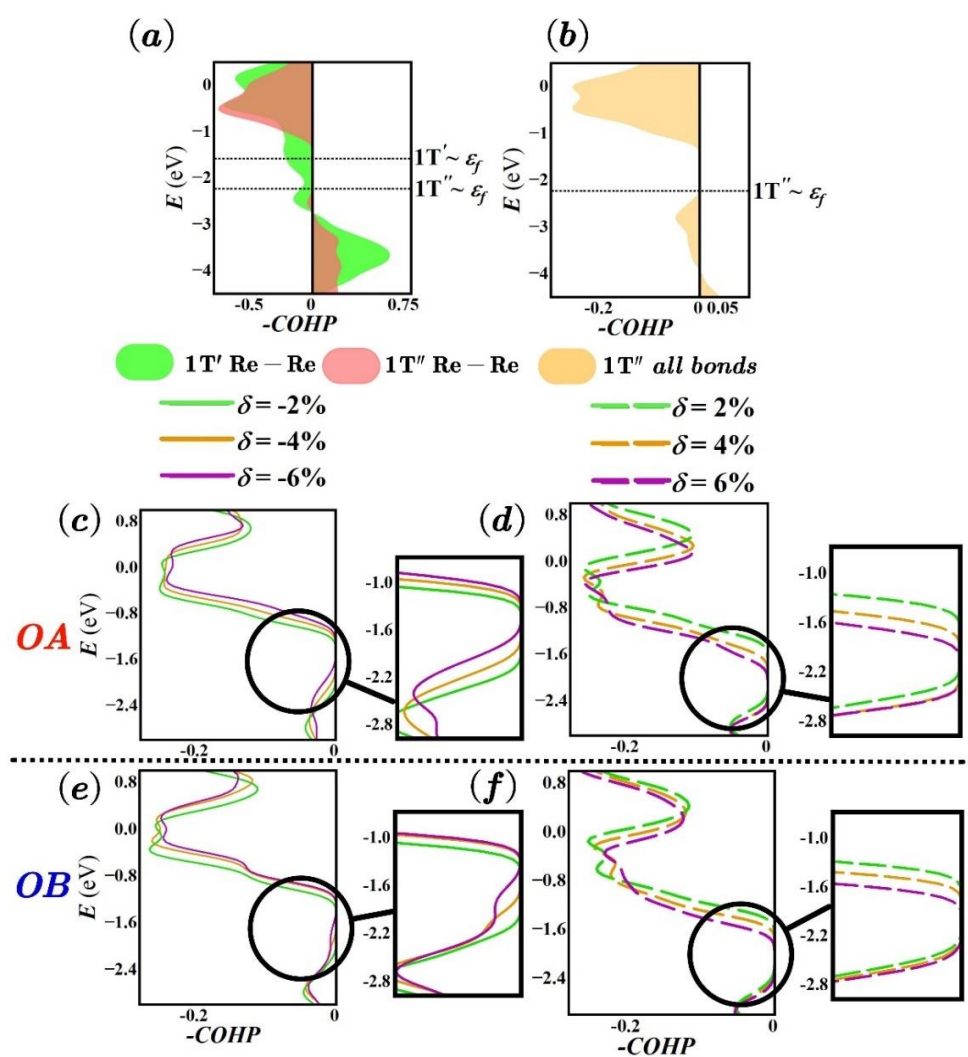

**Fig. 3-8 (a) 1T′/1T″ 相单层 ReSeS 中 Re-Re 成键的 -COHP 曲线，虚线由上到下分别为1T′ 与 1T″ 相的费米能级 (b) 1T″ 相单层 ReSeS 中所有原子成键的 -COHP 曲线与费米能级 (c)-(d) OA 方向与 (e)-(f) OB 方向不同形变下1T″ 相单层 ReSeS 对应的 -COHP 曲线，右侧小图放大了费米能级附近区域**

Fig. 3-8(c)-(f)直观的体现了这一趋势。对于 COHP 来说，成键与反键的强弱体现在两个方面。一是曲线与能量轴围成面积大小的变化，二是相关能量取值本身大小的变化。从计算结果来看，无论施加何种形变，COHP 曲线中峰的形状，所包围面积均无明显改变。故形变对 ReSeS 成键与反键的影响就主要体现在能量的高低关系，特别是 V/CBM 的改变。观察图中黑色圆圈标注区域可以发现，压缩时 VBM 的变化幅度明显大于 CBM；相对应的，拉伸材料会使 CBM 能量迅速下降。该变化的能量尺度一般在 0.3 eV 到 0.6 eV 之间，与上文提到的能带内部色散结构改变相比，造成的能量差异更加明显。所以这才是导致 ReSeS 带隙大小随形变发生改变的更为主要的机制。

最后应该指出，上述通过成键与反键态分析 V/CBM 能量升降的模型存在一个重要的假设，就是我们认为形变下 V/CBM 对应的键形改变是均匀的。这意味着体系中所有的成键与反键将统一的增强或减弱，并且只考虑最终效果。对于一些对称性较好或者成键与反键态在空间中分布有明显轴向的体系，这不是一个好的假设；对于一个体系中所有能量范围的能带，这也不一定是一个好假设，事实上对 ReSeS 中低能区成键态的 COHP 曲线用同样的方法分析就会得出不太好的结果。有时我们需要考虑具体方向形变与具体原子间成键的对应关系，才能得到可靠的结论。但无论如何，这个模型再一次为 ReSeS 费米面附近能量在几何形变下所发生的改变提供了一个较为清晰的物理图像，也启发着我们通过更多角度逐渐逼近 ReSeS 带隙变化的物理全貌，更多讨论会在 4.2 节中进行。

## 3.3 单层 ReSeS 相关电输运性质与光学性质简要介绍

在第三章的最后一节，我们简要对单层 ReSeS 两个与实际应用紧密相连的性质进行了计算。

第一个性质是有效质量，有效质量包括电子有效质量与空穴有效质量，且均在迁移率计算方面有重要意义[42]。有效质量的出发点来自于固体理论中的准经典近似，在这种情况下，电子的动力学行为被近似为波包的行为。通过推导电子在外场下运动的动力学方程，很容易得到电子位于 Γ 点的有效质量 $m_i$ 的表达式，如式 3-2：

$$m_i = \left[ \frac{1}{\hbar^2}\left( \frac{\partial^2 E_C(\bm{k})}{\partial |\bm{k}_i|^2} \right)_{\bm{k}=\bm{k}_0=0} \right]^{-1} \tag{3-2}$$

其中指标 $i$ 代表了倒空间中的不同方向，而 $E_C(\boldsymbol{k})$ 为导带的能带。当取 $E_C(\boldsymbol{k})$ 为价带能谱 $E_V(\boldsymbol{k})$ 时，式 3-2 很自然的代表空穴的有效质量。显然，有效质量的大小取决于能带具体的色散形状，从具体数值关系上看，能带越宽，能带能量随波矢变化便会拥有更大的斜率，进而带来更小的有效质量；从物理图像上看，宽能带意味着波函数程度更大的交叠，进而电子行为趋于非局域，对应更小的有效质量，这与数值关系的分析一致。由于 ReSeS 并不是严格意义的直接带隙半导体，其 VBM 并不位于 Γ 点，我们只计算了 ReSeS 在 Γ 点沿不同方向的电子有效质量，结果如 Fig. 3-9。

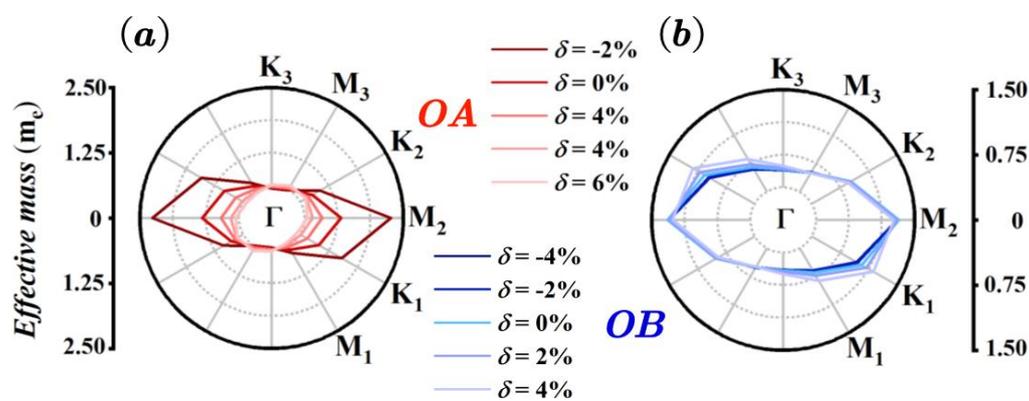

**Fig. 3-9 1T″ 相单层 ReSeS 中 Γ 点沿不同方向的电子有效质量。(a) 为材料沿 OA 轴形变时的情况，(b) 为材料沿 OB 轴形变时的情况**

由结果可知，沿 Γ-$M_2$ 方向 $m_i$ 表现出极大值，意味着波矢沿此方向的电子的运动行为会较为局域。这一特点会随着材料沿 OA 方向发生形变而变得更加明显，当 OA 轴发生仅仅 –2% 的形变时，该方向 $m_i$ 增长约 2.4 倍，有效质量将超过电子质量 $m_e$ 本身，而不同方向 $m_i$ 的极值之比 $m_{i\max}/m_{i\min}$ 更是高达 3.9，体现出极强的各向异性。相比之下，OB 方向的形变并不会导致 $m_i$ 发生显著的变化。这些都说明了 ReSeS 在不同的几何形变之下会拥有差异化的电子动力学行为。进而体现出其在制备电输运相关器件方面蕴含的巨大潜力。

第二个性质是吸收光谱，当然不是上文的红外吸收光谱，而是电子的。材料对光的吸收可以用简单的能级跃迁来理解，当电子吸收了大于带隙能量的光子，便可以跃迁到导带。故吸收光谱的谱线与带隙存在紧密联系。Fig. 3-10 为我们计算所得的数据。

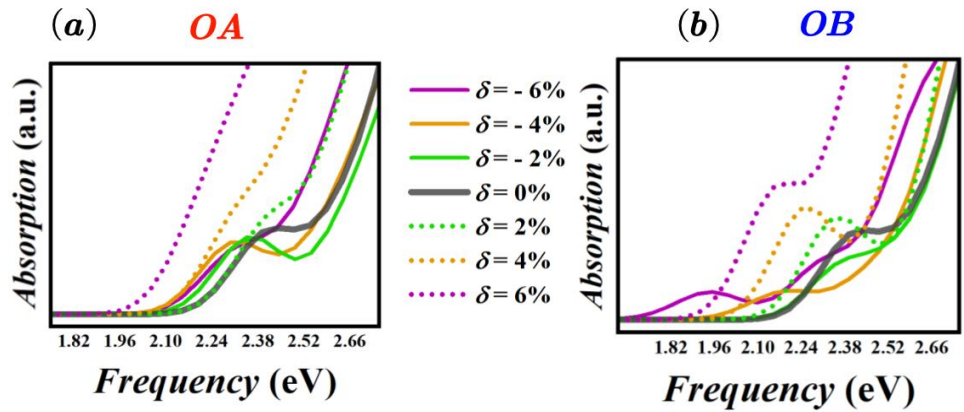

**Fig. 3-10 1T″相单层 ReSeS (a) OA 方向与 (b) OB 方向下不同形变对应的电子吸收光谱**

考虑到不同形变均会导致 ReSeS 带隙缩小，光谱起始峰所对应的能量随着材料形变将会发生蓝移，并且程度越大的形变导致的蓝移越明显。具体而言，OA 轴方向的拉伸形变与 OB 轴方向的压缩形变会造成更为明显的效果，这再一次体现出了 ReSeS 结构具有的各向异性。考虑到其电子光谱的能量范围位于紫外与可见光区域，光谱信息又与带隙信息相对应，带隙信息还与材料形变有关，我们理论上可以通过光学探测的方法得知材料的形变程度，进而对其相关性质进行调控，这又体现出了 ReSeS 参与光学调控的优势。

总的来说，在本节中我们通过计算有效质量与吸收光谱再一次展现了 ReSeS 具有的各向异性，并指出了其参与具体的光电原件设计的可能性。

# 第四章 总结与展望

## 4.1 总结

本文正文整体上分为三个部分：

第一部分我们首先简要介绍了二维材料的发展现状，引出了对于 TMDs 系列材料特性的研究。接着指出 TMDs 材料可能会发生畸变以达到更加稳定的结构。具体而言，我们介绍了$ReS_2/Se_2$在1T′/T″两相相变中发生4聚化这一典型现象。考虑到$ReS_2/Se_2$所具有的低对称性会诱导出新奇的各向异性的物理特性，我们通过组合$ReS_2/Se_2$构建出了具有 Jauns 结构的且对称性更低的ReSeS，并就ReSeS具有的电学性质展开了研究。

第二部分为理论介绍，我们先是主要介绍了本文使用的两种最为主要的理论方法：DFT 与 MLWF 理论。并通过简单的推导明确了二者的基本计算思路。接着我们系统的介绍了本文使用第一性原理计算时所用的具体参数与设置细节。

第三部分为ReSeS电学性质的介绍，并分为了三个小节展开：3.1 节是对ReSeS与$ReS_2/Se_2$基本性质的介绍，内容包括它们的晶格常数，布里渊区与静电势分布。并通过对振动光谱红外拉曼活性的分析体现了ReSeS就$ReS_2/Se_2$发生的对称性破缺。3.2 节我们从ReSeS的能带结构出发，讨论了在几何形变下其带隙发生的变化（在几何形变下带隙缩小），通过分析 V/CBM 能量变化大小指出相比于能带色散形状的改变,能带能量整体上升下降是影响带隙变化的更为主要的原因。接着便由两个模型讨论了其带隙变化的物理图像，在 3.2.1 部分中通过 Wannier 函数思想构造并不紧束缚的解析简化哈密顿量，在这种方法下通过比较 4 个混合轨道的能级差异可以定性的分析出带隙变化规律。但该方法仅适用于对称性差的体系，难以推广，且不具有很好的物理直观；在 3.2.2 部分中通过计算 COHP 分析价带导带的成键与反键特性，并指出是一种"饱和效应"导致了任何形变下带隙均会发生缩小，这个模型在费米面附近得到了较好的结果，并显然有着更为清晰的物理直观。在 3.2.2 部分中，我们还明确了1T′/T″两相相变为一种 Peierls 相变，由此解释了ReSeS带隙的形成。3.3 节中我们计算了ReSeS与具体应用相关的两个性质：电子有效质量与吸收光谱，二者均体现出了ReSeS具有的各向异性，并指出了其在电学，光学器件制备方面具有的潜力与优势。

## 4.2 未来工作展望

本文在 4.2 节中将结合正文工作对三个方向进行展望：

1.对于能带结构变化细节的完善：

3.2 节中提到 ReSeS 中 OB 方向 Re 原子之间存在更强的相互作用，但是并没有指出是什么作用，更没有分析成键与反键，只是简单的说明这会使得 OB 方向的形变对 ReSeS 电子结构造成更明显的改变。麻烦的是，在 3.3 节的分析中，显然 OA 方向的形变对有效质量的改变更为明显。这是为什么呢？具体来说，对于 3.2 节中带隙改变的改变我们更关注能带整体的能量变化，而 3.3 节有效质量则涉及到价带导带与波矢 $k$ 具体的色散关系。这是否意味着几何形变在使得某个方向能带能量变化明显时，削弱了其具体色散结构方面的改变？这是十分有趣的，值得探究。

2.对 ReSeS 构造出一般的有效解析哈密顿量：

这是一个有挑战性的工作，如果我们严格从轨道的角度出发，Re 原子有 5 个 d 轨道，S，Se 原子则分别含有 3 个 p 轨道。这会形成一个不满足任何对称操作的 44×44 的厄密矩阵，分析其中的 hopping 取值大小正负无疑十分繁琐。故我们在 3.2 节中采用了 Re 原子的 s 轨道作为初始猜测，这样可以大大简化模型并得到仅仅 4 个 Wannier 态，但这些态对称性很差，也不是局域的，其形成的原理更像是对 DFT 结果进行强行拟合。不过这在毫无对称操作的 ReSeS 中显得无关紧要，也可以称之为"对称性特权"。从结果来看，它仍然可以作为一种近似模型帮助我们理解带隙变化。然而更理想的是，我们通过某种巧妙的解析方法构建出变量合适的哈密顿量，变量不那么多意味着我们可以进行简单的解析分析；变量不那么少意味着其仍然具有很好的物理直观。在一些对称性较好的体系中，如石墨烯，$MoSe_2$ 与 $WS_2$，这是比较容易实现的[43-44]。但是 ReSeS 对称性太低，二者难以兼顾。虽然困难重重，但是若真的存在这样一个哈密顿量，极有可能优于我们在 3.2 节中提到的两种模型，并使得 ReSeS 带隙缩小的机制更加清晰。

3.通过 Peierls 相变解释 ReSeS 带隙缩小的机制：

在 3.2.2 部分中我们介绍了1T′/T″两相相变的物理图像，但是对其的分析仅仅停留在解释1T″相带隙的形成。一个自然的问题就是：能否通过 Peierls 相变来解释带隙的缩小？我们可以简要的将带隙理解为 d 轨道能带发生的成键-反键劈

裂，故不同形变下的1T′相ReSeS可能对应不同的程度的轨道劈裂，进而影响带隙的大小。Peierls相变往往伴随着电荷密度波（charge density wave，CDW）与费米面嵌套（Fermi surface nesting），一般来说，我们可以通过计算林哈德函数（Lindhard function）估计具体结构中的费米面嵌套波矢[14,45-46]，而研究1T″相ReSeS中 Peierls 相变打开带隙的大小与嵌套波矢的对应关系很可能为我们解释带隙缩小机制提供新的思路。此外，1T′相ReSeS能带结构随几何形变而发生的改变较于1T″相更加明显，所以从这个角度分析还可能提供更为清晰的物理图像。不过需要说明，该想法的可行性并不高，Pasquier 的团队曾指出对于 $t_{2g}$ 轨道半填充的 TMDs，费米面嵌套与体系的不稳定性的相关性较弱[40]。这意味着以上思路可能仍无法把握住ReSeS带隙变化的核心机制，不过仍然可以进行探索与尝试。

总的来说，3.2 节中解释带隙缩小的模型均有缺陷，简化哈密顿量过于粗糙且缺乏物理图像，成键反键模型也只与费米面附近能量变化趋势符合较好，而本节中提出的第 2，3 点，特别是第 2 点，则有可能真正解决这一问题。

## 致谢

在此论文完成之际，我一定要对帮助我的老师，朋友表达衷心的感谢。

感谢指导老师刘力哲教授。

也感谢提供帮助的张海军教授。

也感谢提出宝贵建议的王志俊研究员，吴泉生研究员。

也感谢提供帮助的学长，朋辈：邓浩辰，马俊伟，徐俊奇，王福毅，王俊凯

当然我还要感谢我的亲人，朋友。

祝好！

林天哲

2024 年 5 月 18 日于南京鼓楼南园